\documentclass[12pt]{article}
\pdfoutput=1
\newlength{\abstractwidth}
\usepackage{epsf}
\usepackage{color}
\usepackage{graphicx}

\renewcommand{\thefootnote}{\fnsymbol{footnote}}
\renewcommand{\thanks}[1]{\footnote{#1}}
\newcommand{\starttext}{
\setcounter{footnote}{0}
\renewcommand{\thefootnote}{\arabic{footnote}}}

\newcommand{\bea}{\begin{eqnarray}}
\newcommand{\eea}{\end{eqnarray}}
\newcommand{\ee}{\end{equation}}
\newcommand{\be}{\begin{equation}}
\newcommand{\<}{\langle}
\renewcommand{\>}{\rangle}
\newcommand{\sm}{\smallskip}

\def\cJ{{\cal J}}

\def\cO{{\cal O}}

\def\cT{{\cal T}}

\def\bR{{\bf R}}

\def\det{{\rm det}}

\def\half{ {1\over 2}}
\def\p{\partial}

\def\l({\left(}
\def\r){\right)}

\def\a{\alpha}

\def\ep{\varepsilon}

\def\g{\gamma}

\def\vf{\varphi}
\def\G{\Gamma}
\def\l{\lambda}

\def\g{\gamma}
\def\s{\sigma}

\def\ti{\tilde}

\def\go{g^{(0)}}
\def\gt{g^{(2)}}
\def\gf{g^{(4)}}

\def\gln{g^{({\rm ln})}}
\def\Ao{A^{(0)}}
\def\At{A^{(2)}}

\def\ban{LMN\ }

\def\no{\nonumber}

\def\hti{\ti h}
\def\ati{\ti a}
\def\cti{\ti c}
\def\Hti{\ti H}
\def\Ati{\ti A}

\def\cJh{\hat \cJ}
\def\cTh{\hat \cT}

\begin{document}
\starttext
\setcounter{footnote}{0}

\begin{flushright}
16 May 2011
\end{flushright}

\bigskip

\begin{center}

{\Large \bf Charged Magnetic Brane Correlators \\ and Twisted Virasoro algebras\footnote{This work was
supported in part by NSF grant PHY-07-57702.}}

\vskip .5in

{\large \bf Eric D'Hoker and Per Kraus}

\vskip .2in

{ \sl Department of Physics and Astronomy }\\
{\sl University of California, Los Angeles, CA 90095, USA}\\
{\tt \small dhoker@physics.ucla.edu; pkraus@ucla.edu}

\end{center}

\vskip .2in

\begin{abstract}

Prior work using gauge/gravity duality has established the existence of a quantum critical point
in the phase diagram of 3+1-dimensional gauge theories at finite charge density and background magnetic field.  The critical theory, obtained by tuning the dimensionless charge density to magnetic field ratio, exhibits nontrivial scaling in its thermodynamic properties, and an associated nontrivial dynamical critical exponent.  In the present work, we analytically compute low energy correlation functions in the background of the charged magnetic brane solution to 4+1-dimensional Einstein-Maxwell-Chern-Simons theory, which represents the bulk description of the critical point.  Results are obtained for neutral scalar operators, the stress tensor, and the U(1)-current.
The theory is found to exhibit a twisted Virasoro algebra, constructed from a linear combination of the original stress tensor and chiral U(1)-current.   The effective speed of light in the IR is renormalized downward for one
chirality, but not the other, by finite density, a behavior that is consistent with a Luttinger liquid description of fermions in the lowest Landau level.   The results obtained here do not directly shed
light on the mechanism driving the phase transition, and we comment on why this is so.

\end{abstract}

\newpage

\tableofcontents

\newpage

\section{Introduction and summary of results}
\setcounter{equation}{0}
\label{one}

Holography provides a powerful tool for the study of strongly interacting
fermion systems, including strongly coupled gauge theories. Non-relativistic extensions
to systems at finite temperature, charge density, and external magnetic field,
have applications to physical problems in many areas, including in condensed matter,
heavy ion collisions, and astrophysics \cite{Hartnoll:2009sz,Herzog:2009xv,Sachdev:2010ch}
.
Using holography, thermodynamic functions, correlators, and transport properties
may be calculated in terms solutions to classical gravity equations.
A common setting for many of these systems is provided by RG flows
in 3+1-dimensional gauge theories, with or without supersymmetry.
The gravity dual to a wide class of these gauge theories is governed by
4+1-dimensional Einstein-Maxwell-Chern-Simons theory, whose only
free parameter is the Chern-Simons coupling $k$, which is related to
the number of chiral fermions in the dual gauge theory.

\sm

In the presence of an external magnetic field, the RG flow from 3+1-dimensional
gauge theory in the UV takes us, in the IR, to an effective 1+1-dimensional CFT which is
dominated by massless fermions in the lowest Landau level.  These lowest Landau
level  fermions are only free to move parallel to the magnetic field lines, which
leads to the 1+1-dimensional character of the low energy theory.
This picture is borne out on the gravity side, where
the RG flow is governed by the purely magnetic brane solution which interpolates
between AdS$_5$ in the UV and AdS$_3$ (or BTZ when the temperature is small) in the IR
\cite{D'Hoker:2009mm,Almuhairi:2010rb}.
Holographic calculations of two-point correlators of the U(1) current and stress tensor
further confirm the existence of two Virasoro algebras and a single chiral current algebra
in the IR \cite{D'Hoker:2010hr}. The common central charge of the Virasoro algebras equals
the Brown-Henneaux value \cite{Brown:1986nw} for the AdS$_3$ near horizon geometry
which, via the Cardy formula,
accounts for the specific heat coefficient\footnote{In
the boundary CFT only dimensionless combinations have an invariant physical meaning.
Dividing by a suitable power of the charge density $\rho$, the dimensionless
ratios for the magnetic field $B$, the entropy density $s$, and the temperature $T$ will
respectively be denoted by $\hat B = B/\rho ^{2/3}$, $\hat s = s/\rho$, and $\hat T = T/\rho ^{1/3}$.}
$\hat s /\hat T$  at low $\hat T$.  Such a specific heat is consistent with a
1+1-dimensional Luttinger liquid description \cite{Giamarchi} of the fermionic theory.   The chirality
of the current algebra is given by the sign of $k$, while its level is proportional to $|k|$.
Thus, in this regime, a consistent picture has emerged in which the behavior of
thermodynamic quantities and correlators are found to nicely match.

\sm

As the magnetic field is lowered (or equivalently as the charge density is increased),
the lowest Landau level finds itself
increasingly populated. Eventually, we expect qualitative changes to
occur, both due to occupancy of higher  fermionic Landau levels as well as bosonic
Landau levels, along with any other strong interaction effects at finite density.
Numerical and analytical holographic calculations, carried out in the dual
Einstein-Maxwell-Chern-Simons theory, confirm this expectation and reveal a rich phase
structure as a function of the Chern-Simons coupling $k$ \cite{D'Hoker:2009bc,D'Hoker:2010rz,D'Hoker:2010ij} (see \cite{Lifschytz:2009sz,Jensen:2010vd} for other examples of holographic phase transitions involving magnetic fields and finite density.)  Specifically, for $k > 1/2$, the low
$\hat T$ specific heat coefficient was derived exactly as a function of the magnetic field $\hat B$,
and is given by,
\bea
\label{1a}
{\hat s \over \hat T} = { \pi \over 6} \, { \hat B^3 \over \hat B^3 - \hat B_c ^3} \hskip 1in \hat B > \hat B_c
\eea
While the constancy of $\hat s /\hat T$ as $\hat T \to 0$ is still as in Luttinger liquid theory,
its dependence  on $\hat B$ is not.
As the magnetic field approaches the critical value $\hat B_c$ (which depends only on $k$),
a quantum critical point was discovered at $\hat B = \hat B_c$. The scaling behavior of $\hat s$
at $\hat B = \hat B_c$ was obtained numerically in \cite{D'Hoker:2010rz},
and subsequently established by analytic calculations \cite{D'Hoker:2010ij}, both for $k > 3/4$.
As reviewed in Appendix \ref{appAA}, the picture was completed since by a numerical calculation of the scaling in the range $1/2 < k < 3/4$.
Assembling these data, we have,
\bea
\label{1b}
\hat s \sim \hat T^\alpha \hskip 1in
\alpha = \left \{ \matrix{ {1 \over 3} & & {3 \over 4} \leq k \cr && \cr
{1-k \over k} && \half < k \leq {3 \over 4} \cr} \right .
\eea
This result  signals further strong deviations from canonical Luttinger  liquid behavior.
The full scaling function, accounting for variations in both $\hat T$ and $\hat B$ in the critical region,
was also computed analytically in \cite{D'Hoker:2010ij} for the range $k>1$,
and agrees with the numerical results of  \cite{D'Hoker:2010rz}.

\sm

Remarkably, as discussed in \cite{D'Hoker:2010rz}, the critical exponent 1/3 arises  also in the Hertz/Millis theory \cite{hertz,millis93,Millis02} of
metamagnetic quantum phase transitions, in which a variation in the magnetic field
across a non-zero critical value produces non-analytic behavior in the specific
heat, but no change in symmetry.  Metamagnetic quantum criticality
has been observed \cite{RPMMG} in a wealth
of recent experiments on Strontium Ruthenates, where the entropy density has been mapped out through an extensive region of the temperature and magnetic field parameter space.  Interestingly, as the critical point is approached the specific heat coefficient is found to diverge linearly in $1/(B-B_c)$, just as in (\ref{1a}).

\sm

The purpose of the present paper is dual. On the one hand, we wish to extend
to the case of finite charge density the holographic calculations of various correlators
which have been evaluated in \cite{D'Hoker:2010hr} for vanishing charge density, and understand the
relation with the low $\hat T$ thermodynamics already established for finite charge density.
On the other hand, if the holographic calculations of thermodynamics are really
relevant to metamagnetic quantum phase transitions, then obtaining the corresponding
holographic correlators should allow one to address transport problems in
metamagnetic materials close to quantum criticality in a model with only a single
free parameter $k$. The results of this paper
will constitute a modest first step in both directions, and will  be summarized below.

\subsection{Summary of results}

In this paper, 2-point correlators will be evaluated in the IR regime for a neutral scalar, for the U(1)
current, and for the stress tensor, in the presence of the charged magnetic brane solution.
The method of overlapping expansions may be justified for small momenta, and will be used throughout.

\sm

Calculating the 2-point correlators of a neutral scalar in the IR regime, we identify effective left- and right-moving
excitations with different propagation speeds. While left-movers propagate
at the speed of light (with respect to the standard boundary metric) as they do for
the purely magnetic brane, the speed of the right-movers is less than the speed of light
in the presence of non-zero charge density. We compare these holographic predictions
for the dynamics of chiral excitations with those of Luttinger liquid theory \cite{Giamarchi}, in which the
speed is also predicted to be renormalized downward. Finally, we argue in favor of a relation between
the entropy density scaling of (\ref{1b}) for $1/2<k<3/4$ and the dynamical scaling properties
of the scalar two-point functions. We leave unexplained, however, the precise nature of the crossover
to the critical exponent 1/3, and the precise value of the coupling $k$ at which this cross-over should occur.

\sm

Calculating the 2-point functions of the U(1) current $\cJ_\pm $ and the stress tensor $\cT_{\pm \pm}$
reveals  that their IR behavior is governed by two {\sl Virasoro algebras, one of which is twisted with the
chiral current algebra}. The positive chirality correlators are given by the following expressions, valid for
all $k > 1/2$,
\bea
\label{1c}
\left \< \cJ_+ (x^+) \, \cJ_+ (y^+) \right \> & \sim & - { kc \over 2 \pi^2} \, { 1 \over (x^+-y^+)^2}
\no \\
\left \< \cJ_+ (x^+) \, \cT_{++} (y^+) \right \> & \sim &  + { kc \mu  \over 2 \pi^2} \, { 1 \over (x^+-y^+)^2}
\no \\
\left \< \cT_{++} (x^+) \, \cT_{++} (y^+) \right \> & \sim &  - { kc \mu^2 \over 2 \pi^2 } \, { 1 \over (x^+-y^+)^2}
+ { c \over 8 \pi^2} \, { 1 \over (x^+-y^+)^4}
\eea
where $c$ is the Brown-Henneaux central charge, and $\mu$ is the chemical potential.
The non-vanishing negative chirality correlator is given by,
\bea
\label{1d}
\left \< \cT_{--} (x^-) \, \cT_{--} (y^-) \right \> & \sim &  { c \over 8 \pi^2} \, { 1 \over (x^- -y^-)^4}
\eea
These correlators have higher order IR corrections, whose quantitative evaluation would
require going beyond the overlapping expansion method. Qualitatively, however, we
know that some of these higher order corrections arise from double and higher trace
operators which are generated by the RG flow towards the IR \cite{Heemskerk:2010hk,Faulkner:2010jy}, just as they were shown
to arise in the purely magnetic case in \cite{D'Hoker:2010hr}. Upon forming the combination
\bea
\label{1e}
\cT_{++} ^{(0)} = \cT_{++} + \mu \cJ_+
\eea
the chiral current algebra of $\cJ_+$ decouples from the Virasoro algebra of $\cT^{(0)}_{++}$,
and we are left with two decoupled Virasoro algebras, and one decoupled chiral current algebra,
just as we had in the case of the purely magnetic case. These results are valid for all $k > 1/2$,
and constitute perhaps the most important result of this paper. The net effect of turning on
a non-zero charge density (or equivalently, a non-zero chemical potential $\mu$) is
simply to {\sl twist} the positive chirality Virasoro algebra by the admixture of $\mu \cJ_+$, as well as renormalizing the speed of propagation.
These results confirm the presence of a unique underlying CFT for all values of the magnetic field,
but offer no immediate insight into the dynamical scaling properties of the entropy density in (\ref{1b}).

\subsection{Organization}

The remainder of this paper is organized as follows. In section \ref{two}, we review the
Einstein-Maxwell-Chern-Simons theory and the salient properties of its charged magnetic brane solutions.
In section \ref{three}, we calculate the 2-point correlators of a neutral scalar field of arbitrary mass,
and discuss some of their immediate physical consequences in section \ref{fourA}. In section \ref{four},
we outline the general structure of the 2-point correlators of the U(1) current and the stress tensor.
In section \ref{five} we set up the technical calculation of the linear fluctuations in the
gauge field and metric using overlapping expansions in light-cone gauge for the near region
to LMN and Fefferman-Graham gauge in the far region. The detailed calculations of
the twisted Virasoro terms in the current and stress tensor 2-point functions are
derived in sections \ref{six} and \ref{seven}. A brief discussion of some open questions is presented in section \ref{eight}.  In Appendix \ref{appAA} we review the behavior of the low temperature specific heat exponent as a function of $k$.  In Appendix \ref{appA}, perturbative and
WKB approximations to the scalar correlators are developed. In Appendix \ref{appB},
the full linearized fluctuation equations around the charged magnetic brane solution
are derived in light-cone gauge, while the gauge transformations required in
the overlap region are constructed in Appendix \ref{appC}.

\newpage

\section{Review of the charged magnetic brane solution}
\setcounter{equation}{0}
\label{two}

The action of five-dimensional Einstein-Maxwell-Chern-Simons theory with a
negative cosmological constant and Chern-Simons coupling $k$ was given
in \cite{D'Hoker:2010hr}. Here, we will
need the corresponding Bianchi identity $dF=0$ and field equations,
\bea
\label{20a}
0 & = &  d  \star  F + k F \wedge F
\no \\
R_{MN} & =  &
4  g_{MN} +{1 \over 3} F^{PQ} F_{PQ} g_{MN} -2 F_{MP} F_N{}^P
\eea
Throughout, we will take $k \geq 0$, since a sign reversal of $k$ is equivalent
to a parity transformation.
For the supersymmetric value $k=2/\sqrt{3}$,  this system is a consistent
truncation for all supersymmetric compactifications of Type IIB or M-theory
to AdS$_5$ (see \cite{Buchel:2006gb,Gauntlett:2006ai,Gauntlett:2007ma}).
Solutions of (\ref{20a})  are guaranteed to be solutions of the full 10 or 11
dimensional supergravity field equations, and to provide holographic duals to the infinite
class of supersymmetric field theories dual to these more general
supersymmetric AdS$_5$ compactifications.

\subsection{Boundary stress tensor and current}

We will be considering asymptotically AdS$_5$ solutions, for which the metric
and gauge field admit a Fefferman-Graham expansion \cite{FG}.
Introducing a radial coordinate $\rho$, defined
such that the AdS$_5$ boundary is located at $\rho=\infty$,
the metric takes the asymptotic form
\bea
\label{20b}
ds^2 & = & {d\rho^2 \over 4\rho^2} +g_{\mu\nu}(\rho,x) dx^\mu dx^\nu
\no \\
g_{\mu\nu}(\rho,x) & = & \rho \go_{\mu\nu}(x) + \gt_{\mu\nu}(x)
+ {1 \over \rho}\gf_{\mu\nu}(x)
+{\ln \rho \over \rho}  \gln_{\mu\nu}(x)  + \cdots
\eea
while the gauge field takes the asymptotic form,
\bea
\label{20c}
A & = & A_\mu(\rho,x) dx^\mu
\no \\
A_\mu(\rho,x)& = & \Ao_\mu(x)+{1\over \rho} \At_{\mu}(x) +\cdots
\eea
Here $x^M = (\rho,x^\mu)$, with $\mu = +,-,1,2$. The coefficients $\gt_{\mu\nu}$,
$\gln_{\mu\nu}$, and the trace of $\gf_{\mu \nu}$ are fixed by the Einstein equations
to be local functionals of the conformal boundary metric $\go_{\mu\nu}$.

The boundary stress tensor and current
\cite{Balasubramanian:1999re,deHaroSkenderis}
are defined in terms of the variation of the on-shell action
with respect to $\go_{\mu \nu}$ and $\Ao_\mu$ respectively (see \cite{D'Hoker:2010hr}).
In terms of the Fefferman-Graham data the result is,
\bea
\label{20d}
4\pi G_5 T_{\mu\nu}(x) & = & \gf_{\mu\nu}(x) +{\rm local}
\no \\
2\pi G_5 J_\mu(x) & = & \At_\mu(x)+{\rm local}
\eea
Indices are raised and lowered using the conformal boundary metric
$\go_{\mu\nu}$.  The local terms denote tensors constructed locally
from $\go_{\mu\nu}$ and $\Ao_\mu$.  In this paper, as in
\cite{D'Hoker:2010hr}, we are interested in computing two-point correlation
functions of operators at non-coincident points, to which the local terms
in (\ref{20d}) do not
contribute. Henceforth, we will drop all local terms.

\subsection{Ansatz, Reduced field equations, and symmetries}

The charged magnetic brane solution was constructed in \cite{D'Hoker:2010ij}.
It belongs to a class of solutions characterized by a constant background magnetic
field, rotation invariance
around this magnetic field, and translation invariance in all boundary directions
$x^\mu$. The Ansatz is conveniently expressed in the so-called ``LMN" gauge,
\bea
\label{21a}
ds^2 & = & { dr^2 \over L^2-MN} + M(dx^+)^2 + 2 L dx^+ dx^- + N (dx^-)^2
+ e^{2V} dx^idx^i
\no \\
A & = & b x^1 dx^2 + A_+dx^+ + A_- dx^-
\eea
The index $i=1,2$ labels the coordinates of the plane orthogonal to the magnetic field.
The magnetic field strength $b=F_{12}$ is constant, and by rescaling $x^{1,2}$, may
be set  to a convenient value, $b =\sqrt{3}$. The functions $L,M,N,V,A_\pm$ depend
only upon the holographic coordinate~$r$.
The reduced field equations were given in \cite{D'Hoker:2010ij},
with the identifications $E=A_+', P=-A_-'$,
\bea
\label{21aa}
{\rm M1} &&\quad  \left((NE+LP)e^{2V}\right)'+2kbP =0
\\ \no
{\rm M2} && \quad    \left((LE+MP)e^{2V}\right)'-2kbE =0
\\ \no
{\rm E1} && \quad L''+2V'L' +4(V'' +V'^2)L -4PE=0
\\ \no
{\rm E2} && \quad M'' +2V'M'+4(V'' +V'^2)M +4E^2=0
\\ \no
{\rm E3} && \quad N''+2V'N'+4(V'' +V'^2)N+4P^2  =0
\\ \no
{\rm E4}  && \quad  f (V')^2 +  f' V'  -6 +{1 \over 4}(L')^2 - {1 \over 4} M'N' + b^2 e^{-4V}
+ MP^2 + 2 LEP + NE^2 =0
\\ \no
{\rm fV} && \quad f'' + 4 f'V'+ 2 fV''+ 4 f (V')^2 - 24 =0
\eea
Throughout, primes will denote derivatives with respect to the holographic coordinate,
and we will use the abbreviation,
\bea
\label{21bb}
f = L^2 -MN
\eea
Notice that  LMN gauge of (\ref{21a}) differs from Fefferman-Graham gauge
of (\ref{20c}) only in the structure of the term proportional to $dr^2$.
Finally, under constant $SL(2,\bR)$ transformations of the coordinates~$x^\pm$,
\bea
\label{21cc}
\left ( \matrix{ \ti x^+ \cr \ti x^- } \right ) = \Lambda ^{-1} \left ( \matrix{ x^+ \cr x^- } \right )
\hskip 1in \Lambda \in SL(2,R)
\eea
the Ansatz (\ref{21a}) is invariant provided the fields transform as follows,
\bea
\label{21dd}
\left ( \matrix{ \ti A_+ \cr \ti A_- } \right ) = \Lambda ^t
\left ( \matrix{ A_+ \cr A_- } \right )
\hskip 1in
\left ( \matrix{ \ti M & \ti L \cr \ti L & \ti N } \right ) = \Lambda ^t
\left ( \matrix{ M & L \cr L & N } \right ) \Lambda
\eea
while the field $V$ and the combination $f$ are invariant.

\subsection{Charged magnetic brane solution}

The basic charged magnetic brane solution satisfies the consistent restriction,
\bea
\label{21b}
N=P=0
\eea
As a result, the functions $L=L_0$ and $V=V_0$ satisfy the reduced field equations
E1, E4, and fV of the purely magnetic background solution \cite{D'Hoker:2010hr}.
The E4 equation is a constraint whose derivative
vanishes by virtue of  equations E1 and fV. Although no analytical solution is
known, the relevant parameters governing the small and large $r$ asymptotics
are available numerically, and this will suffice for our purpose. The remaining functions
$A_+=A_0$ and $M=M_0$ were solved for in \cite{D'Hoker:2010ij}, and are given by,
\bea
\label{21d}
A_0 (r) & = & {c_V c_E \over kb} \, e^{2kb \psi (r)}
\no \\
M_0(r) & = & L_0(r) \left ( - { \a_\infty \over 2b}
- 4kb \int ^r _\infty { dr' A_0 (r')^2 \over L_0(r')^2 \, e^{2V_0(r')}}  \right )
\eea
where $\psi$ is defined by,
\bea
\label{21e}
\psi (r) =  \int ^r _\infty { dr' \over L_0(r') \, e^{2V_0(r')}}
\eea
The parameters $c_V, c_E$, and $\a_\infty$ entering these functions specify
the asymptotic behavior of these functions, as we will now spell out.

\subsection{Asymptotic AdS$_5$ behavior}

As $r \to \infty$, the solution approaches AdS$_5$, and we have the following
asymptotics,
\bea
\label{23a}
L_0 (r) & = & 2 (r-r_0) - { 3 \ln r \over c_V^2 \, r} + \cO(r^{-1})
\no \\
e^{2V_0(r)} & = & c_V (r-r_0) + \cO(r^{-1})
\no \\
A_0 (r) & = & {c_V c_E \over kb} - {c_E \over r} + \cO(r^{-2})
\no \\
M_0(r) & = & - { \a_\infty \over b} \, r + \cO(r^0)
\no \\
\psi (r) & = &  - { 1 \over 2 c_V \, r} + \cO(r^{-2})
\eea
The interpretation of these constants was given in \cite{D'Hoker:2010ij}.
The constants $c_V$ and $r_0$  characterize the purely magnetic solution.
The asymptotics $E_0 (r) \sim c_E/r^2$ of the electric field $E_0=A_0'$
shows that $c_E$ is proportional to the electric charge density $\rho$ of the
dual state, while the asymptotics of $A_0$ gives its chemical
potential  $\mu$.  The precise relations are as follows,
\bea
\rho = 4 c_E \, \left ( { 2 b \over  \a _\infty} \right )^\half
\hskip 1in
\mu  = { c_V c_E \over kb}
\eea
where the coefficient $\alpha_\infty$ arises from the asymptotics of the metric function $M_0$.

\subsection{Asymptotic near-horizon behavior}

As $r \to 0$, the solution approaches a family of exact solutions for which
$V$ is constant,
\bea
\label{23b}
F & = & b dx^1 \wedge dx^2 + q \, r^{k-1} dr \wedge  dx^+
\no \\
ds^2 & = & {dr^2 \over 4 b^2}
- \left ( \a_0 r + { 2 q^2 r^{2k} \over k (2k-1)} \right ) (dx^+)^2
+ 4br dx^+ dx^- + dx^i dx^i
\eea
The corresponding functions are given by,
\bea
\label{23c}
L_0(r) & = & 2br + \cO(r^{1 + \sigma}) \hskip 1in \sigma = - \half + {\sqrt{57} \over 6}
\no \\
e^{2V_0(r)} & = & 1 + 2 r^\sigma + \cO(r^{2 \sigma})
\no \\
A_0 (r) & = & { q \over k} \, r^k + \cO( r^{k +\sigma})
\no \\
M_0(r) & = & - \a_0 r + \cO(r^{1 +\sigma})
\no \\
\psi (r) & = &  {\ln r \over 2b} + \psi _0 + \cO(r^\sigma)
\eea
The constant $\psi_0$ is a property of the purely magnetic solution only.
On the other hand, $q$ and $\a_0$ are associated with the presence of
non-zero charge density.

The metric appearing in the $(r, x^+, x^-)$ directions is the same as a 2+1-dimensional Schrodinger solution \cite{Son:2008ye,Balasubramanian:2008dm}, or alternatively a null-warped solution \cite{Anninos:2010pm}.

\subsection{Relations between asymptotic parameters and regularity}

The key asymptotic data in the large and small $r$ asymptotics are
related as follows,
\bea
\label{23e}
q & = & {c_V c_E \over b} \, e^{2kb\psi_0}
\no \\
\a_0 & = & \a_\infty - 16 c_V^2 c_E^2 J(k) \hskip 1in
J(k) = { 1 \over 2k} \int ^\infty _0 { dr \, e^{4kb \psi (r) } \over L_0(r)^2 e^{2V_0(r)}}
\eea
They link the large $r$ and small $r$ behavior of $A_0$ and $M_0$. The integral
$J(k)$ is convergent for $k > 1/2$, and is positive, so that we must have
$\a_0 < \a _\infty$ as long as $q\not=0$.

\sm

Regularity of the solution requires $M_0(r) \leq 0$ for all $r>0$.footnote{More precisely, if this condition is not obeyed then
the zero temperature solutions studied here cannot be obtained as limits of
smooth, finite temperature solutions.}
When $q \not= 0$, the near-horizon metric (\ref{23b}) is regular only when
$k > 1/2$,  a condition we will assume throughout.
Since $M_0(r) /L_0(r) + \a_\infty /(2b)$ vanishes at $r=\infty$, and is
monotonically increasing for decreasing $r$, regularity also requires
 $0 \leq \a_0$. The relation between $\a_\infty$
and $\a_0$ was shown in~\cite{D'Hoker:2010ij}\footnote{The coefficients
$\a_\infty$ and $\a_0$ were denoted respectively by $\a$ and $\ti \a$
in~\cite{D'Hoker:2010ij}.}
to be governed by the (normalized) magnetic field $\hat B = B / \rho ^{2/3}$,
where $B=2b/c_V$,
\bea
\label{23f}
{ \a_0 \over \a_\infty } = 1- { \hat B _c^3 \over \hat B^3}
\hskip 1in
\hat B _c^3  = {12 J(k) \over c_V}
\eea
Thus, regularity of the charged magnetic brane solution forces $\hat B \geq \hat B_c$,
with the critical limit $\hat B = \hat B_c$ corresponding to a quantum phase
transition \cite{D'Hoker:2010ij}.  For $\hat B < \hat B_c$ a new branch of
solutions opens up; these were studied numerically in \cite{D'Hoker:2010ij}, but no analytical results are currently available.

\subsection{The action of $SL(2,\bR)$}
\label{twoseven}

$SL(2,\bR)$ transformations which preserve the consistent restriction
$P=N=0$ of (\ref{21b}) may be parametrized as follows,
\bea
\label{27a}
\Lambda =
\left ( \matrix{ \lambda_0 & 0 \cr  - \lambda _0 \lambda & \lambda_0^{-1} \cr  } \right )
\eea
They will map charged magnetic brane solutions with different parameters into
one another. Denoting the transformed quantities by tildes,
as in (\ref{21dd}), we find,
\bea
\label{27b}
\tilde q & = & \lambda _0 \, q
\no \\
\tilde \a _0 & = & \lambda _0^2 ( \a_0 + 4b \lambda)
\no \\
\tilde \a _\infty & = & \lambda _0^2 ( \a_\infty + 4 b \lambda)
\eea
We see that the parameter $\lambda_0$ rescales the charge $q$, while
the parameter $\lambda$ changes the normalized magnetic field.
In particular, all charged magnetic brane solutions may be obtained as
$SL(2,\bR)$ maps of the critical solution for which $\a_0=0$,
$\hat B = \hat B_c$, and $\a_\infty = \a_c= 16c_V^2 c_E^2 J(k)$,
the magnetic field being given by,
\bea
\label{27c}
1- {\hat B _c ^3 \over \hat B^3} = {4b\lambda \over \a_c + 4 b \lambda}
\eea
Turning on a non-zero temperature, as we did in \cite{D'Hoker:2010ij},
or non-zero momenta, as we will do here, will make these solutions
physically inequivalent, as the $SL(2,\bR)$ transformation will now also
change the temperature and/or the momenta. The practical advantage
is that, without loss of generality, we can restrict to studying the
solution corresponding to the critical magnetic field $\hat B = \hat B_c$,
since the solution for other (finite) values of $\hat B$ may be derived
from the critical solution by applying the above $SL(2,\bR)$ transformation.
Thus, henceforth, we set $\a_0=0$, and $\a_\infty = \a_c$.

\newpage

\section{Correlators of a neutral scalar field}
\setcounter{equation}{0}
\label{three}

In the case of the purely magnetic background solution, the fluctuations of the
gauge field decoupled from the metric fluctuations to linearized order, so that
the correlator between the $U(1)$ current and the stress tensor vanished \cite{D'Hoker:2010hr}.
For the charged magnetic brane solution, no such decoupling takes
place, a fact that renders the problem more involved. A simpler warm-up
case is provided, however, by studying the fluctuations of a free neutral
scalar field in the presence of the charged magnetic brane solution.
Although the neutral scalar will only probe the metric of the background
solution, and not its gauge field, its study will provide valuable information
for scalar correlators that would be much more difficult to obtain for the
full stress tensor and current correlators.

\subsection{Reduced Field Equation}

We consider a free neutral scalar field $\Phi$ with mass $m$. It obeys
the wave equation,
\bea
\label{30a}
{ 1 \over \sqrt{g}} \p_M \left ( \sqrt{g} g^{MN} \p_N \Phi \right ) - m^2 \Phi =0
\eea
The background being invariant under translations in $x^\mu$,
and equation (\ref{30a}) being linear, it will suffice to consider plane wave solutions
with momenta $p_\mu = (p_+, p_-, p_1, p_2)$. Since we will be
interested only in the $x^\pm$-dependence, we set $p_1=p_2=0$,
which leaves,
\bea
\label{30b}
\Phi (r, x^\pm) = \phi (r, p_\pm) \, e^{ipx}
\eea
where we continue to use the notations of \cite{D'Hoker:2010hr}, namely,
$px = p_+ x^+ + p_- x^-$, and $p^2 = p_+ p_-$. In the background of the
charged magnetic brane solution, the wave equation (\ref{30a}) reduces to,
\bea
\label{30c}
L^2 \phi '' +{1 \over K} (KL^2)' \phi ' - {2 p^2 \over L} \phi
+ {p_-^2 M \over L^2} - m^2 \phi =0
\eea
Throughout, we will often use the abbreviation $e^{2V}=K$.
Since the metric is unperturbed in this system, we have dropped the
subscript 0 on the functions $L,M,K$. No analytic solution is known to
the reduced scalar equation (\ref{30c}), which comes as no surprise since
the functions $K,L,M$ themselves are not known analytically. Nonetheless,
for small momenta (precise conditions will be spelled out later),
we may use the method of overlapping expansions which was used
already in  \cite{D'Hoker:2010hr} to calculate the entropy density
of the charged magnetic brane solution for low temperature, as a function of $\hat B$.
Following this method, we solve (\ref{30c}) in the low momentum limit,
by matching the solutions in overlapping near and far regions,
corresponding to small and large $r$ respectively. As announced in
section \ref{twoseven}, we restrict to analyzing the critical solution
corresponding to $\hat B = \hat B_c$, since all other solutions may be
recovered by the $SL(2,\bR)$ transformations of section \ref{twoseven}.

\subsection{Solutions in the far region}

In the far region, momentum dependence in (\ref{30c}) is to be neglected,
which requires $ p^2 \ll r$ and $p_-^2 \ll r$. We want the far region to extend
down to $r\ll 1$ so that the far region will overlap non-trivially with the near region;
this requires $p^2 \ll 1$ and $p_-^2 \ll 1$. For $k<1$, there is, however,
an additional requirement: since $M(r) \sim r^{2k}$ for $r \ll 1$, the above overlap
condition will hold provided we have $p_-^2  r^{2k-2} \ll 1$ as well (this
condition is automatic for $k>1$ in view of the requirement $p_-^2 \ll 1$ already
imposed earlier).
Note that it is possible to satisfy all these conditions by making $p_-$ small,
and yet allowing $p_+$ to be of order 1.

\sm

Assuming that the above conditions are satisfied,
all momentum dependence in (\ref{30c}) may be dropped, and we
are left with the reduced equation in the far region,
\bea
\label{31a}
L^2 \phi '' +{1 \over K} (KL^2)' \phi '  - m^2 \phi =0
\eea
which involves only the data of the purely magnetic background solution.
Clearly, no analytical solution for $\phi$ is available, since $K$ and $L$
themselves are known only numerically. The data we need from the far
region, however, depend on only a small number of parameters which may
be computed numerically, if so desired.

\sm

To see how this works, consider
(\ref{31a}) for $r \gg 1$, where $L(r)\sim 2r$ and $K(r) \sim c_V r$,
\bea
\label{31b}
4 r^2 \phi '' + 12 r \phi ' - m^2 \phi =0
\eea
For $r \gg 1$, the general solution is given by,
\bea
\label{31c}
\phi (r) = a_+ r^{\nu_+} + a _- r^{\nu_-} \hskip 1in
\nu _\pm = -1 \pm \sqrt{1 + m^2/4}
\eea
For $r \ll1 $, but still in the far region where momenta are dropped,
we have $L(r)\sim 2 br$ and $K(r) \sim 1$, so that (\ref{31a}) becomes,
\bea
\label{31d}
12 r^2 \phi ''+ 24 r \phi ' - m^2 \phi =0
\eea
The general solution of (\ref{31d}) is
\bea
\label{31e}
\phi (r) = b_+ r^{+{\nu \over 2} - \half} + b_- r^{-{\nu \over 2} - \half}
\hskip 1in \nu =  \sqrt{ 1 + m^2/3}
\eea
By linearity of (\ref{31a}), there exists a linear relation between
the coefficients $a_\pm$ and $b_\pm$, which we may express as follows,
\bea
\label{31f}
\left ( \matrix{ a_+ \cr a_- \cr} \right ) = S \left ( \matrix{ b_+ \cr b_- \cr} \right )
\hskip 1in
S = \left ( \matrix{ S_{++} & S_{+-}  \cr S_{-+} & S_{--} } \right )
\eea
The correlator of the scalar field $\Phi$ is given by the ratio,
\bea
\label{31g}
G = { a_- \over a_+} = { S_{-+} + S_{--} (b_-/b_+)
\over S_{++}  + S_{+-} (b_-/b_+)   }
\eea
The matrix $S$ depends only on the data of the purely magnetic solution,
and on the mass $m$ of the scalar, but not on any momenta.
The ratio $b_-/b_+$ will be governed by the dynamics in the near region,
and will depend  on the momenta $p_\pm$.

\subsection{Solutions in the near region: purely magnetic case}

We begin by solving for the behavior of the scalar in the purely magnetic
case, since this may be done analytically and simply.
The near region is defined by $r \ll 1$, so that $L(r) \sim 2br$ and $K(r) \sim 1$.
Equation (\ref{30c}) then takes the form,
\bea
\label{32a}
12 r^2 \phi '' + 24 r \phi ' - { p^2 \over br} \phi - m^2 \phi=0
\eea
By the change of variables $r = z^{-2}$, and the redefinition of the function
$\phi (r) = z \ti \phi (z) $, equation (\ref{32a})  is transformed into a modified
Bessel equation,
\bea
\label{32b}
z^2 \ti \phi '' + z \ti \phi ' - \left ( {p^2 \over b^3} z^2 + \nu^2 \right ) \ti \phi =0
\eea
Of its two linearly independent solutions, $K_\nu$ vanishes at the horizon
$z = \infty$, while $I_\nu$ diverges there.  Retaining only the regular solution,
and expressing it in terms of $\phi(r)$, we obtain,
\bea
\label{32c}
\phi(r) = { 2 \sin \nu \pi \over \pi \sqrt{r} } K_\nu \left ( \sqrt{ p^2 \over b^3r} \right )
\eea
In the $r \gg p^2$ part of the near region, $\phi$ is given by the following approximate
behavior,
\bea
\label{32d}
\phi (r) \sim
{ 1 \over \G (1-\nu) } \left ( {p^2 \over 4 b^3 } \right ) ^{- {\nu \over 2} }
r^{+{\nu \over 2} - \half}
-{ 1 \over \G (1+\nu) } \left ( {p^2 \over 4 b^3 } \right ) ^{+ {\nu \over 2} }
r^{-{\nu \over 2} - \half}
\eea
This formula was derived using the familiar definition,
\bea
\label{32dd}
K_\nu(z) & = & {\pi \over 2 \sin (\pi \nu)} \left ( I_{-\nu}(z) - I_\nu(z) \right )
\eea
as well as the $z \to 0$ asymptotics of $I_\nu (z)$, given in (\ref{A4}) of Appendix B.

\subsubsection{IR behavior of 2-point correlator: purely magnetic case}

The far and near regions overlap when $p^2 \ll r \ll 1$, so that the solution (\ref{31e})
valid in the far region must match the solution (\ref{32d}) in the near region.
We see that their functional dependence on $r$ indeed coincides, which
allows us to extract the key ratio,
\bea
\label{32e}
{ b_- \over b_+} = - {\G (1-\nu) \over \G (1+\nu) } \left ( {p^2 \over 4 b^3 } \right ) ^\nu
\eea
Assuming a fixed value of $\nu$ away from the positive integers,
and that $p^2 \ll 1$, we find  $b_-/b_+ \ll 1$. The scalar two-point function
of (\ref{31g}) may then be expanded in powers of $b_-/b_+$, and the
leading non-trivial momentum dependence identified,
\bea
\label{32f}
G(p) & = & {S_{-+} \over S_{++}} +   { \det (S)  \over (S_{++})^2 } \, C_\nu  p^{2 \nu}
+ \cO(p^{4 \nu})
\no \\
C_\nu  & = & - { \G (1-\nu) \over  \G (1+\nu) ~ 2^{2 \nu} b^{3\nu}}
\eea
The first term in $G(p)$ produces a delta function in position space, which is local,
and may be dropped. The determinant factor $\det (S)$ is proportional
to the Wronskian of the differential equation (\ref{31a}), evaluates to
$\det (S)= 3 \nu (c_V)^{-1} (\nu _+ - \nu _-)^{-1}$, and thus never vanishes.
As a result, the leading IR behavior of the scalar Green function
is given by $p^{2 \nu}$.

\subsection{Solutions in the near region: charged magnetic brane case}

In the near region $r \ll 1$ we have $K \sim 1$,  $L \sim 2br$, and $M$ is given by
\bea
\label{35a}
M(r) = - a_M r^{2k} \hskip 1in a_M =  { 2q^2 \over k(2k-1)}
\eea
where we are restricting to the critical solution and set $\a_0=0$.
All momentum dependence in (\ref{30c}) now needs to be retained,
resulting in the equation,
\bea
\label{35b}
12 r^2 \phi '' + 24 r \phi ' - { p^2 \over br} \phi - {p_-^2 a_M \over 12 } r^{2k-2} \phi
- m^2 \phi =0
\eea
Changing variables to $r=1/z^2$, and redefining  $\phi (r) = z \ti \phi (z)$,
transforms (\ref{35b}) into,
\bea
\label{35c}
z^2 \ti \phi '' + z \ti \phi ' - \left ( { p^2 \over b^3} z^2 + \nu^2 + { p_-^2 a_M \over 36}
z^{4-4k} \right ) \ti \phi =0
\eea
Here, $\nu$ is given by (\ref{31e}). For $q=a_M=0$, this equation coincides with (\ref{32b}), which has already been solved earlier. For $p_+=0$, but $p_- \not= 0$,
equation (\ref{35c}) may be transformed into a
modified Bessel equation, and solved analytically as well, for all $k$.

\sm

Henceforth, we will assume $q, p_\pm \not=0$. Rescaling $z$ and the
scalar field as follows,
\bea
\label{35d}
z= {b^{3/2} x  \over  p}
\hskip 1in
\ti \phi (z) = \vf (x)
\eea
transforms  (\ref{35c}) into,
\bea
\label{35e}
x^2 \vf '' + x \vf ' - \left ( x^2 + \nu ^2 + \xi ^2 x^{4-4k} \right ) \vf =0
\eea
Thus, for fixed $k$ and $\nu$, and up to the above rescaling
of $z$, the scalar field equation in the near region
intrinsically depends only on a single combination $\xi$  of the momenta, given by,
\bea
\label{35f}
\xi = \xi_0 (p_+)^{k-1} (p_-)^k
\hskip 1in
\xi _0 = q \left ( {3^{1-3k}   \over 2k(2k-1) }  \right ) ^\half
\eea
Let $\vf (x)$ denote the solution to (\ref{35e}) that is smooth
at the horizon, $x \to \infty$. In the overlap region, we have $p^2 \ll r \ll 1$
which translates to $p \ll x \ll 1$ in terms of the coordinate $x$,
and the function $\vf (x)$  takes the simplified  form,
\bea
\label{35g}
\vf (x) \sim c_-(\xi) x^\nu + c_+ (\xi) x^{-\nu}
\eea
The dependence on $k$ and $\nu$ of coefficients $c_\pm (\xi)$ has not been
exhibited here. Transforming back to the holographic coordinate $r$, we have,
\bea
\label{35h}
\phi  (r) \sim c_-(\xi) \left ( {p \over b^{3/2}} \right )^\nu r^{-{\nu \over 2} - \half}
+ c_+(\xi) \left ( {p \over b^{3/2}} \right )^{-\nu} r^{{\nu \over 2} - \half}
\eea

\subsubsection{IR behavior of 2-point correlator: general form}

In the overlap region, where we have $p^2 \ll r \ll 1$, the functional form (\ref{35h}) of
$\phi(r)$ in the near region coincides with the functional form
(\ref{31e}) in the far region. Matching the corresponding coefficients gives,
\bea
\label{38a}
{ b_- \over b_+} = C_\nu (\xi) p^{2 \nu}
\hskip 1in
C_\nu (\xi) =  { c_-(\xi)  \over c_+ (\xi)  \, b^{3 \nu}}
\eea
Again, since $p^2 \ll 1$ we have  $b_-/b_+ \ll 1$, so that the two-point
correlator for the scalar may be reliably approximated as follows,
\bea
\label{38b}
G(p) = {S_{-+} \over S_{++}} +  { \det (S) \over  (S_{++})^2 } \,
C_\nu ( \xi) p^{2 \nu} + \cO(p^{4 \nu})
\eea
The  $\xi \to 0$ limit reduces to the purely magnetic case so that we have
$C_\nu (0) = C_\nu$ with $C_\nu$ given in (\ref{32f}). Also, the $p_+ \to 0$
limit, while keeping $p_-$ fixed, should be smooth. For $k >1$, this limit implies
$\xi \to 0$, and again leads to the purely magnetic case. For $k<1$, the
$p_+ \to 0$ limit implies $\xi \to \infty$, and smoothness of $G$ then requires,
\bea
\label{38c}
C_\nu(\xi) \, \sim \, \xi ^{ { \nu \over 1-k}} \hskip 1in \hbox{as} \qquad \xi \to \infty
\eea
The precise coefficient can be computed analytically but will not be needed here.
The remaining $p_-$ dependence of $b_-/b_+$ is then of the form $(p_-)^{\nu/(1-k)}$,
as may also be seen directly from scaling arguments on the original
equation (\ref{35c}) with $p^2=p_+=0$.

\sm

Beyond these asymptotic limits, it does not appear possible to solve
equation (\ref{35e}) for the near region behavior of the scalar field
analytically for generic value of $k$. Besides the trivial $k=1$ case,
there is one special value $k=3/4$ where a complete solution may
be obtained in terms of Whittaker functions.

\subsection{The solvable special case of $k=3/4$}

For the special value $k=3/4$, equation (\ref{35e}) is related to
the Whittaker equation \cite{AS}.  The solution to (\ref{35e}) which vanishes
for large $x$ is given by,
\bea
\label{36a}
\vf (x) = e^{-x} x^\nu U \left ( \half + \nu + {\xi^2 \over 2}, 1 + 2 \nu, 2x \right )
\eea
where $U$ is the confluent hypergeometric function in the notation
of \cite{AS}. In the overlap region, its behavior simplifies to,
\bea
\label{36b}
\vf (x) \sim { x^\nu \over \G (1+2 \nu) \G ( \half - \nu + {\xi^2 \over 2} )}
-  { x^{-\nu} \over \G (1-2 \nu) \G ( \half + \nu + {\xi ^2 \over 2} )}
\eea
where we have omitted an immaterial multiplicative constant common to
both terms on the right hand side of (\ref{36b}). In terms of the holographic
coordinate $r$, we find,
\bea
\label{36c}
\phi (r) \sim
{ (2p b^{-3/2} )^\nu \, r^{-\half -{\nu \over 2}} \over
   \G (1+2 \nu) \G ( \half - \nu + {\xi^2 \over 2} )}
-  { (2p b^{-3/2} ) ^{-\nu} \, r^{-\half + { \nu \over 2}} \over
\G (1-2 \nu) \G ( \half + \nu + {\xi ^2 \over 2} )}
\eea
As a result, we find the following formula,
\bea
\label{36d}
{b_- \over b_+} = C_\nu (\xi) \, p^{2 \nu}
\hskip 1in
C_\nu (\xi) =  - { \G (1-2\nu) \G \left ( \half + \nu + {\xi^2 \over 2} \right ) \over
\G (1+2\nu) \G \left ( \half - \nu + {\xi ^2 \over 2} \right )} \left ( {4 \over b^3} \right )^\nu
\eea
One may check that the $\xi \to 0$ limit of the above expression reproduces
the general result $C_\nu (0) = C_\nu$ of (\ref{32f}), by using the
duplication formula for the $\G$-function.

\sm

There are three more
special cases, $k=5/8$, $k=7/8$, and $k=3/2$ where the solutions to
the differential equation  (\ref{35e}) are expressible in terms of Heun functions.
As of yet, we have found insufficient information on the asymptotic
behavior of these functions to make use of them in the generation
of the scalar two-point correlator for these values of $k$.

\subsection{Comparing expansions with the exact  $k=3/4$ solution}

Beyond obtaining the exact near-horizon solution for  $k=3/4$ it is
possible to derive the functional form of the expansion
in powers of $\xi$ of $C_\nu (\xi)$, through a combination of
perturbative and WKB methods. These calculations are
carried out in Appendix \ref{appA}, and their results may be
summarized as follows. The perturbative expansion in $\xi$
is valid when $\xi \ll 1$, while the WKB expansion holds when $\hbar \ll 1$,
\bea
\label{39a}
\half  <  k < 1 & \quad \hbox{in powers of} \quad &
\xi^2 \sim {p_- ^{2k} \over p_+ ^{2-2k}}
\no \\
\half < k < 1 & \quad \hbox{in powers of} \quad &
\hbar \sim   \left ( {  p_+ ^{1-k} \over  p_- ^k} \right ) ^{1/(2k-1)}
\eea
Both expansions may be applied to the $k=3/4$ special case, where we have an
exact solution. For $k=3/4$, the above expansion parameters become,
\bea
\label{39b}
\xi ^2 \sim {p_- ^{3/2} \over p_+ ^{1/2}}
\hskip 1in
\hbar = {1 \over \xi^2}
\eea
The exact solution of (\ref{36d})  for $k=3/4$,
\bea
\label{39c}
C_\nu (\xi) \sim {\G \left ( \half + { \xi ^2\over 2} + \nu \right )
\over
\G \left ( \half + { \xi ^2 \over 2}  - \nu \right )}
\eea
clearly admits a Taylor series expansion in powers of $\xi^2$ for small $\xi$.
It admits an expansion for large $\xi$, which may be computed
using the following series expansion of the $\G$-function,
\bea
\label{39d}
{\G (y+ 2 \nu) \over \G(y)} = y^{2 \nu} \left ( 1 + {\nu (2 \nu -1)  \over y}  + \cO(y^{-2}) \right )
\hskip 1in y = \half +{\xi^2 \over 2} - \nu
\eea
showing that the expansion parameter is $\hbar \sim 1/\xi^2$ for large $\xi$.
This result matches the prediction of WKB given in (\ref{39a}), and (\ref{39b}).

\sm

Note that, contrarily to the case of the current correlators in the purely magnetic brane
solution, the corrections in powers of $\xi$, either for large $\xi$ or for small $\xi$,
do not readily admit an interpretation in terms of double trace deformations of
the action and operators, because the expansions in powers of $\xi$ or $1/\xi$ are
not given by  geometric series. It would be interesting to investigate whether
the structure seen here can be accounted for by multi-trace operator perturbations,
or requires a novel paradigm.

\sm

We close with a discussion of the case $ k > 1$. The combination $\xi ^2$ now vanishes
uniformly as $p_\pm \to 0$.  It is multiplied by $z^{4-4k}$, a factor which becomes singular at $z=0$
when $k > 1$. This singularity is beyond the domain of validity of the near-horizon
region, in which we are considering equation (\ref{35e}). Thus, it is more suitable to
start from equation (\ref{35b}), with $r \ll 1$ and to carry out a power
expansion in $p_-^2 a_M$. The lowest order contribution to
the series is just the purely magnetic solution.


\newpage

\section{Physical Applications}
\setcounter{equation}{0}
\label{fourA}

In this section we discuss two immediate physical applications of the
calculations of scalar correlators, given in the previous section. A first
concerns the connection between the critical scaling law of the entropy density
of (\ref{1b}). A second deals with the velocities of the chiral excitation
modes, their dependence on the magnetic field, and similar effects encountered in
Luttinger liquid theory.

\subsection{Dynamical scaling exponent and entropy scaling form}
\label{dynam}

We begin by  commenting on the dynamical scaling exponent
that emerges in the near-horizon geometry, and its relation with
the thermodynamic scaling relation (\ref{1b}) for the entropy density.
The metric of the near-horizon geometry at the critical point is of the
Schr\"odinger type and given by,
\bea
\label{37a}
ds^2 = { dr^2 \over 4 r^2} - { 2 q^2 r^{2k} \over k(2k-1)}  (dx^+)^2 + 4 b r dx^+ dx^- + dx^i dx^i
\eea
As shown already in \cite{D'Hoker:2010ij}, it is invariant under the following scale transformations,
\bea
\label{37b}
r \to \lambda r \hskip 1in x^+ \to \lambda ^{-k} x^+ \hskip 1in x^- \to \lambda ^{k-1} x^-
\eea
This dynamical scaling symmetry is reflected in the IR behavior of the correlators
of the scalar field through the dependence on only the combination,
\bea
\label{37c}
\xi \sim (p_-)^k (p_+)^{k-1}
\eea
which is invariant under the scalings of (\ref{37b}). Reverting to coordinates
$x ^+ \to t$ and $x^- \to x_3$, and frequency/momentum $p_+ \to \omega$ and $p_- \to \kappa$,
used in \cite{D'Hoker:2010ij} to study thermodynamics, we
readily identify the standard dynamical scaling exponent, defined by $\omega \sim \kappa ^z$, as
\bea
\label{37d}
z = { k \over 1-k}
\eea
Scaling arguments alone then predict the scaling relation for the entropy density $\hat s$
as a function of $\hat T$ in $d$ space-dimensions, $\hat s \sim \hat T^{d/z}$.
Assuming that the effective IR theory is indeed a 1+1-dimensional CFT, as well
as the value of the dynamical scaling $z$ of (\ref{37d}) would lead to
\bea
\label{37e}
\hat s \sim \hat T ^\alpha \hskip 1in \alpha = {1-k \over k}
\eea
Comparison with (\ref{1b}) reveals agreement with the full gravity calculation
only for $1/2 < k < 3/4$, but disagreement for $k > 3/4$. Thus, the key problem in
explaining (\ref{1b}) becomes to understand the transition across $k=3/4$.
The near-horizon expansion in powers of $\xi$ will also result when we will
study the 2-point correlators of the current and stress tensor.

\subsection{Speed of chiral excitations}

The scalar correlator computed in the background of the charged magnetic brane
may be used to calculate the velocities at which the low momentum scalar excitations
propagate in the dual CFT. For $k >1$, the low momentum limit $|p_\pm | \ll 1$ uniformly
corresponds to $ \xi \ll 1$. For $1/2 < k < 1$ the limit is not uniform, however,  and
we need to distinguish two regimes, loosely characterized by $|p_-| \ll |p_+| $ and $|p_-| \gg |p_+|$,
and more precisely specified by  $\xi \ll 1$ and $\xi \gg 1$ respectively. Up to local terms, and up to an
overall constant, the Green function of (\ref{38b}) in both regimes is given as follows,
\bea
\label{37f}
\xi \ll 1 & \hskip 1in & G(p) \sim (p_+p_-)^\nu
\no \\
\xi \gg 1 && G(p) \sim (p_-)^{\nu \over 1-k}
\left ( 1 + c_1 (p_+)^{{1-k \over 2k-1}} (p_-)^{- { k \over 2k-1}} + \cO(\hbar^2) \right )
\eea
Since the second line arises only in the interval $1/2<k<1$, it is instructive to specialize
to the solvable value $k=3/4$, where $G(p)$ is given by,
\bea
\label{37g}
G(p) \sim (p_-)^{4\nu} + c_1 (p_+)^{\half} (p_-)^{4 \nu - {3 \over 2}} + \cdots
\eea
Taking the Fourier transform gives the propagator as follows.
For $1/2<k<1$ and $x^+$ small compared to $x^-$, as well as for $k>1$ and
both $x^\pm$ large, we have,
\bea
\label{37h}
G(x^\pm) \sim { 1 \over (x^+)^{\nu+1} (x^-)^{\nu+1} }
\eea
For $1/2<k<1$ and $x^+$ large compared to $x^-$, we have,
\bea
\label{37i}
G(x^\pm) \sim { 1 \over (x^+)^{{3 \over 2} } (x^-)^{4 \nu - \half} }
\eea
where the leading $(p_-)^{4 \nu}$ term in (\ref{37g}) is local and may be dropped.

\sm

As is manifest from the factorized form of these contributions, the propagation of
left-movers and right-movers are independent from one another. To calculate the
velocity of propagation of each chiral mode, we need to focus on the normalization
of the boundary metric which, for the critical solution $\hat B = \hat B_c$, is given by,
\bea
ds^2 = 4 dx^+ dx^- + \beta (dx^+)^2 + dx^i dx^i
\hskip 1in \beta = - { \alpha _c \over b}
\eea
Thus we see that the propagators of (\ref{37h}) and (\ref{37i}) were in fact
expressed in an oblique coordinate system. To find the physical velocities of
propagation, we need to convert to a set of standard coordinates, $\tilde x^\pm$, defined by
\bea
\tilde x^+ = x^+ \hskip 1in \tilde x^- = x^- + \beta x^+
\eea
in which the boundary metric is given by,
\bea
ds^2 = 4 d\tilde x^+ d \tilde x^- + dx^i dx^i
\eea
Introducing cartesian coordinates $t,x_3$ by $\tilde x^\pm =(x_3 \pm t)/2$,
the speed of light is set to 1. The propagation of the chiral modes follows
the singularities of the propagator. For the mode $x^+=0$, we have $\tilde x^+= x_3+t=0$,
corresponding to a left-moving chiral excitation propagates at the speed of light.
For the mode $x^-=0$, we have $\tilde x^- - \beta \tilde x^+=0$, or
$(1-\beta) x_3 = (1 + \beta ) t$, which corresponds to a right-moving chiral excitation
propagating at the speed
\bea
v = { 1+ \beta \over 1- \beta}
\eea
Since $\beta <0$,  the right-mover propagate slower than light.
We may obtain a formula for the velocity in terms of intrinsic scale invariant
physical observables in the boundary CFT by expressing $\alpha_\infty$
as follows, $\alpha _\infty = 4c_E^2 c_V^3 \hat B^3/3$, using the formulas of
(\ref{23e}) and (\ref{23f}). Using next the expression for the chemical potential
$\mu$ and its scale invariant form $\hat \mu = \mu/ \rho ^{1/3}$, we further
express $c_V c_E$ in terms of $\mu$, resulting in $\alpha _\infty = 8 b k^2 \hat \mu^2 \hat B^2$,
and the velocity,
\bea
v = { 1 - 2 k^2 \hat \mu^2 \hat B^2 \over 1 + 2 k^2 \hat \mu^2 \hat B^2}
\eea

\subsection{Comparison with Luttinger liquid theory}

Luttinger liquid theory extends the Fermi liquid picture in 1+1 dimensions
by including all (marginal) four-Fermi interactions. We begin by reviewing the
Luttinger liquid construction. As a finite charge density is turned on,
a Fermi surface develops, which in 1+1 dimensions consists of just
two points with opposite momenta $\pm k_F$, as depicted in Figure 1.

\begin{figure}[h]
\begin{centering}
\includegraphics[scale=0.6]{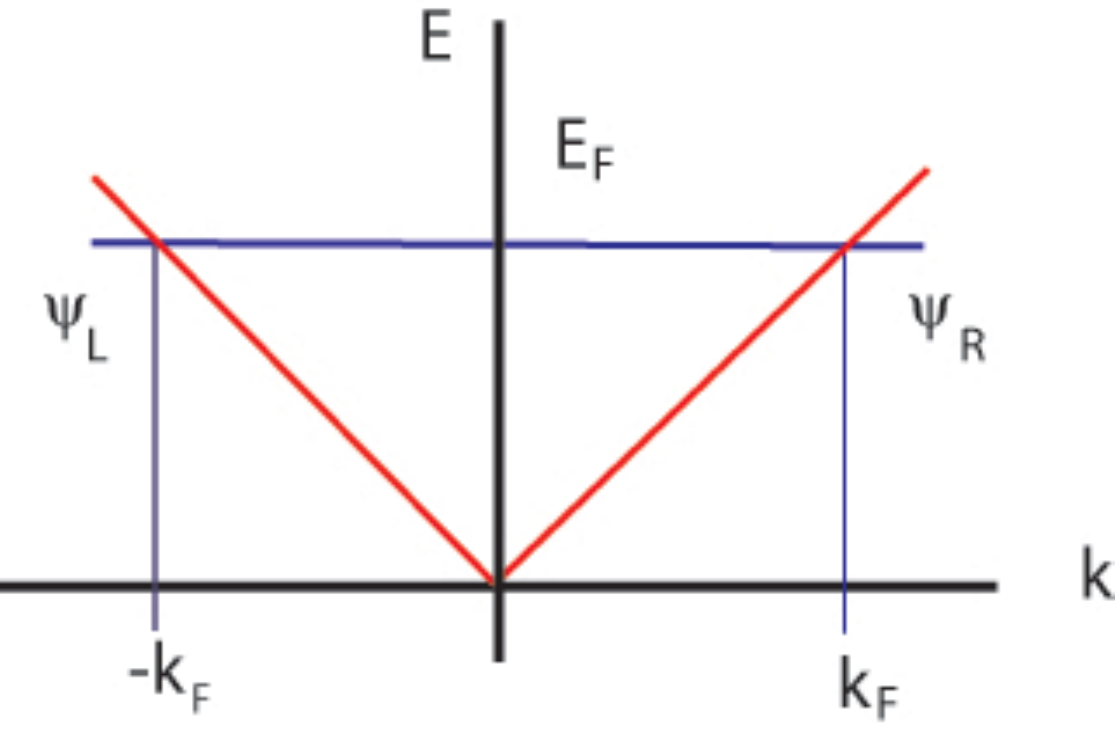}
\caption{Fermi ``surface" and 1+1-dimensional chiral fermions $\psi _L$ and $\psi _R$}
\label{fig1}
\end{centering}
\end{figure}

At those points, we have chiral fermionic degrees of freedom. Assuming the presence of
just a single species, (spinless fermions) we may describe these with the complex fields
$\psi _L$ and $\psi_R$. Their Hamiltonian contains the standard free part,
\bea
H_0 = v_F \int dx \left ( \psi ^\dagger _L (i \p_x + k_F) \psi _L +  \psi ^\dagger _R (i \p_x - k_F) \psi _R \right )
\eea
where $v_F$ is the Fermi velocity.
In Luttinger liquid theory, all (marginal) four-Fermi  interactions are included in the Hamiltonian as well,
\bea
H_{{\rm int}} = 2 \pi v_F \int dx \left ( g_2 \left ( \psi _L ^\dagger \psi _L \right ) \left ( \psi _R ^\dagger \psi _R \right )
+  g_4 \left ( \psi _L ^\dagger \psi _L \right ) ^2 +  g_4 \left ( \psi _R ^\dagger \psi _R \right ) ^2 \right )
\eea
where $g_2$ and $g_4$ are dimensionless couplings. The Luttinger model considered here
is left-right symmetric, but not Lorentz invariant. It may be solved by bosonization
to a single free bosonic excitation, whose velocity is renormalized to,
\bea
v = v_F \sqrt{ (1+g_4)^2 - g_2^2 }
\eea
Qualitatively, this effect is similar to the change in velocity we have found for
the scalar excitations in the presence of the charged magnetic brane. There,
however, the renormalization of velocity affects only one chirality, not both.
The renormalization of velocity also affects the entropy density dependence on $T$,
as the specific heat coefficient for the above Luttinger liquid is given by,
\bea
{s \over T} = { \pi \over 3 v}
\eea
Comparing this expression with the entropy density for the charged magnetic brane,
we see that a decreased velocity leads to an increase in specific heat coefficient,
an effect that is qualitatively consistent with the gravity picture.

\newpage

\section{Structure of current and stress tensor correlators}
\setcounter{equation}{0}
\label{four}

In this section, we will derive the general structure of the 2-point functions
involving the U(1) current $\cJ$ and the stress tensor $\cT$ in the state dual
to the charged magnetic brane solution. We will show that the
correlators may be organized in terms of a Taylor expansion in powers
of the charge density parameter $q$ of the background solution (defined in (\ref{23e}) and related
formulas). This general structure will be confirmed by explicit calculations
in subsequent sections.

\subsection{Review of  correlators in the purely magnetic solution}

The 2-point correlators of $\cJ$ and $\cT$ in the state dual to the purely magnetic
background solution (for which $q=0$) were evaluated to leading
order in the approximation of small momenta \cite{D'Hoker:2010hr}.
The results are given as follows,\footnote{The properly normalized
2-dimensional current and stress tensor operators, which were denoted
by $\cJh$ and $\cTh$ in~\cite{D'Hoker:2010hr},
(see equations (4.23) and  (6.18) of \cite{D'Hoker:2010hr} for their precise
definition) will be denoted respectively by $\cJ$ and $\cT$ respectively here,
i.e.  without hats.}
\bea
\label{91a}
\left \< \cJ_+ (p) \, \cJ_+ (-p) \right \> \Big |_{q=0}
& \sim & {kc \over 2 \pi} \, {p_+ \over p_-}
\no \\
\left \< \cT_{++} (p) \, \cT_{++} (-p) \right \> \Big |_{q=0}
& \sim & {c \over 48 \pi} \, {p_+^3 \over p_-}
\no \\
\left \< \cT_{--} (p) \, \cT_{--} (-p) \right \> \Big |_{q=0}
& \sim & {c \over 48 \pi} \, {p_-^3 \over p_+}
\eea
All other correlators vanish for $q=0$ in this approximation. In fact, since the
purely magnetic background solution is invariant under charge conjugation,
the mixed correlators $\< \cJ \, \cT \>$ in the dual state vanish exactly. The
overall strength of the correlators is governed by,
\bea
\label{91aa}
c= {b \over 2G_3}
\eea
which is the Brown-Henneaux central charge for AdS$_3$ with radius $\ell_3 = b^{-1}$.

\subsection{Extended symmetries for the charged magnetic brane solution}

As $q$ is turned on, the charge conjugation and 2-dimensional Lorentz symmetries
of the purely magnetic background are spontaneously broken by the non-zero charge
density $q$. Charge conjugation reverses the sign of $q$, while a Lorentz
boost transforms $q$ as a (light-like) Lorentz vector. The latter motivates
the notation,
\bea
\label{91b}
q=q_+
\eea
where the subscript + stands for a Lorentz index.  {\sl Extended transformations},
which transform the fields as well as the symmetry breaking parameter,  will leave the
background solution invariant, as always  with spontaneous symmetry breaking.
Concretely, transforming all Lorentz vector and tensor fields as well as the parameter
$q_+$ under {\sl extended Lorentz transformations} leaves the charged magnetic
brane solution invariant. Similarly, the solution is invariant under {\sl extended
charge conjugation}. In the dual theory, these extended Lorentz and charge conjugation
symmetries will be realized through Ward identities for the corresponding
spontaneously broken symmetries. It is this invariance that we will use to
organize the $q_+$-dependence of all correlators for the charged magnetic
brane solution.

\sm

Extended charge conjugation symmetry allows the correlators $\< \cJ \, \cT \>$
to be non-vanishing functions which are odd in $q_+$ and permits
$q_+$-dependent  corrections to the correlators $\< \cJ \, \cJ \>$ and
$\< \cT \, \cT \>$ which are even in $q_+$. Extended Lorentz invariance
further restricts the correlators
as follows. In momentum space, the correlators may be organized
through their dependence on the combinations $q_+ p_-$
and $p^2=p_+ p_-$ which are invariant under extended Lorentz symmetry,
and respectively odd and even under extended charge conjugation.
(In particular, these are the only combinations that will enter the full
reduced field equations (\ref{B5}), as will be discussed later.)

\subsection{The approximation}
\label{approx}

The full analytic calculation of the correlators for general values of momenta is at present out of reach, and
we will instead construct an approximate solution with the help of overlapping
expansions.  Our aim is to extract the leading behavior of certain correlators in the limit of low momenta.    It will be helpful to spell out here precisely what this approximation entails.
In momentum space, the approximation consists in neglecting the following contributions:
\begin{enumerate}
\item purely local terms produced by polynomials in $p_+$ and $p_-$;
\item non-integer powers of $p^2$, such as  $p^{4k}$
and $p^{4 \sigma +2}$ which arise from the near region solution;
\item positive integer powers of $p^2$ which arise from the far region solution;
\item positive integer powers of $q_+p_-$ which arise from the far region solution for $q_+ \not=0$.
\end{enumerate}
Any term which is suppressed by a contribution of the above type relative to a leading term will be neglected.
As we will see, these rules will allow us to isolate the leading long-distance behavior of certain correlators, and will amount to setting other correlators to zero.
Within this approximation, the correlators of (\ref{91a}) in the purely magnetic
background solution were derived in \cite{D'Hoker:2010hr} by explicit calculation
using the overlapping expansion methods.
The corrections to (\ref{91a}) referred to in point 4. are absent here, while the corrections of
point 3. are immaterial since their presence would produce only purely local terms.

\subsection{General functional dependence on $p_\pm$ and $q$, and analyticity}

Extended Lorentz symmetry restricts each 2-point correlator to be of the form,
\bea
\label{92a}
(p_+)^{s_+} (p_-)^{s_-}F(q_+p_-, p_+p_-)
\eea
where $s_\pm $ are non-negative integers, $s_+ - s_-$ is the total spin of the
correlator, and $F$ is a function of the extended Lorentz invariants
$q_+p_-$ and $p^2=p_+ p_-$. Extended charge conjugation symmetry
prohibits fractional powers of $q_+$, since these would produce non-trivial
monodromy under $q_+ \to - q_+$.
The $q_+ \to 0$ limit will be assumed to be smooth, since it physically
corresponds to letting the charge density tend to zero, which is a smooth process.
As a result, $F$ must have a Taylor series expansion in powers of $(q_+p_-)$.

\sm

In view of point 2. of the approximations made in section \ref{approx}, we are to
omit all fractional powers of $p^2$, leaving us to retain only integer powers of $p^2$.
In view of points 1. and 3. of section \ref{approx}, and the fact that $s_\pm \geq 0$,
we may omit all non-negative integer powers of $p^2$.
Putting all together, we find that $F$ must take the form,
\bea
\label{92b}
F(q_+p_-, p_+ p_-) = \sum _{m=0} ^\infty \, \sum _{n > 0 }
F_{m,n} (q_+p_-)^m (p_+p_-)^{-n}
\eea
The sum over $m$ is restricted to even (resp. odd) integers for
correlators that are even (resp. odd) under extended charge conjugation.
Finally, we retain only the leading long distance contribution, corresponding to the term in (\ref{92b}) with smallest $m$ and largest $n$.

\subsection{The trace of the stress tensor}
\label{trace}

The trace $\cT$ of the full 4-dimensional stress tensor $\cT_{\mu \nu}$ is
governed by the trace anomaly and is thus a local function of the metric and
gauge field. As a result, any 2-point correlator involving $\cT$ is local and,
within the approximations spelled out in section \ref{approx}, effectively vanishes,
a result we will denote by $\cT \approx 0$.
Concretely, $\cT$ is given by,\footnote{Here, and throughout,
we set $\cT_{11}=\cT_{22}=\cT_K$, and $\beta = - \a _c /(4b)$
with $\a_c$ given in (\ref{23a}). By construction, the holographic stress tensor
is symmetric, so that we have $\cT_{+-}=\cT_{-+}$.}
\bea
\label{91ca}
\cT = g^{\mu \nu} \cT_{\mu \nu} = 4 \cT_{+-} - 4 \beta \cT_{--} + {8 \over c_V} \cT_K
\eea
where we have used the form of the conformal boundary metric $g_{\mu \nu}$
for the charged magnetic brane in the oblique coordinate system (\ref{21a}),
\bea
\label{91c}
ds^2 _\infty = dx^+ dx^- + \beta (dx^+)^2 + {c_V \over 4} dx^i dx^i
\eea
Furthermore, it will be proven in section \ref{six2} that the 2-point correlator
of $\cT_K$ with any one of the operators $\cJ_\pm$, $\cT_{\pm \pm}$, $\cT_{+-}$, and
$\cT_K$ vanishes within the approximations of section \ref{approx}, so that we also have
$\cT_K \approx 0$. Combining this result with $\cT \approx 0$ and  (\ref{91ca}),
we find that
\bea
\label{91cc}
\cT_{+-} -  \beta \cT_{--} \approx 0
\eea
since its 2-point correlator with any one of the operators $\cJ_\pm$, $\cT_{\pm \pm}$,
$\cT_{+-}$, and $\cT_K$ vanishes within the approximations of section \ref{approx}.
Result (\ref{91cc}) should come as no surprise: putting the metric $ds_\infty^2$ in
standard form by changing coordinates to,
\bea
\label{91e}
\ti x^+ = x^+
\hskip 1in
\ti x^- = x^- + \beta x^+
\eea
condition (\ref{91cc}) translates to the customary vanishing trace condition $\tilde \cT _{+-} \approx 0$.

\sm

The trace condition (\ref{91cc}) leads to relations between the correlators for various
components of the stress tensor. Some care is needed, however, in keeping these
relations consistent with the approximations of section \ref{approx}. For example, to order
$q^0$ the correlators of $\cT_{--}$ with $\cT_{+-}$ and $\cT_{++}$ vanish, and thus it follows
from (\ref{91cc}) that the correlators of $\cT_{+-}$ with $\cT_{+-}$ and $\cT_{++}$ vanish
at least to order $q^2$, all within the approximations of section \ref{approx}. The correlator of
$\cT_{+-} $ with $\cT_{--}$ to order $q^2$ is proportional to $\beta p_-^3/p_+$.
Contrarily to naive appearances, this correlator vanishes within the approximation of section \ref{approx}.
Indeed, since we have $\beta = - \a_c/b \sim q^2$, the correlator exhibits a factor of $(q_+p_-)^2$,
and must be set to zero. Analogous arguments hold for the correlators of
$\cT_{+-}$ with the currents, and lead to their vanishing. Collecting the results obtained in this way, we have,
\bea
\label{91n}
\< \cT_{+-} (p) \cT_{\pm \pm} (-p) \> & \sim & 0
\no \\
\< \cT_{+-} (p) \cT_{+-} (-p) \> & \sim & 0
\no \\
\< \cT_{+-} (p) \, \cJ_\pm \, (-p) \> & \sim & 0
\eea
Higher order corrections in $q_+$ will be accompanied by higher powers in $p_-$ and are
to be neglected as well. Thus (\ref{91n}), and thus the operator relation $\cT_{+-} \approx 0$,
will be valid to all orders in $q_+$ within the approximations of section \ref{approx}.

\subsection{The current component $\cJ_-$}
\label{current}

It will be shown by explicit calculation  in section \ref{JT}
that the following relation holds in all 2-point correlators within the approximations
of the overlapping expansion method,
\bea
\label{93a}
\cJ_- - Q_+ \cT_{--} \approx 0
\hskip 1in
Q_+ = {k \a_c \over  c_V c_E}
\eea
This relation is analogous to (\ref{91cc}). As a result, the fate of the 2-point correlators
involving $\cJ_-$, under the approximations of section \ref{approx}, is analogous to the
fate of the correlators involving the $\cT_{+-}$ component of the stress tensor, and we deduce the relations,
\bea
\label{93b}
\< \cJ_- (p) \, \cJ_\pm (-p) \, \> & \sim & 0
\no \\
\< \cJ_- (p) \cT_{\pm \pm}  (-p) \> & \sim & 0
\no \\
\< \cJ_- (p) \cT_{+-}  (-p) \> & \sim & 0
\eea
Equivalently, the operator relation $\cJ_- \approx 0$ will be valid to all orders in $q_+$,
within the approximations of section \ref{approx}.

\subsection{Conservation of current and stress tensor}
\label{cons}

Gauge invariance and translation invariance in $x^\mu$ guarantee that the
current $\cJ_\mu $ and the stress tensor $\cT_{\mu \nu}$ must be conserved,\footnote{More precisely, the divergence of the current is nonzero in general due to the chiral anomaly, but since the anomaly is purely local, it may be set to zero consistently within the present framework.}
\bea
\label{97a}
\p_+ \cJ_- \, + \, \p_- \cJ_+ \, - 2 \beta \, \p_- \cJ_- ~ & = & 0
\no \\
\p_+ \cT_{-+} + \p_- \cT_{++} - 2 \beta \, \p_- \cT_{-+} & = & 0
\no \\
\p_+ \cT_{--} + \p_- \cT_{+-} - 2 \beta \, \p_- \cT_{--} & = & 0
\eea
Throughout we set to zero the momenta $p_i=0$, so that
no derivates $\p_i$  occur and $\cT_K$ does not enter.
In view of the relations $\cT_{+-} \approx 0$ and $\cJ_- \approx 0$ derived in
the preceding two subsection, these conservation relations reduce to {\sl chiral
conservation} equations,
\bea
\label{97b}
\p_- \, \cJ_+ \,  & \approx & 0
\no \\
\p_- \cT_{++}  & \approx & 0
\no \\
\p_+ \cT_{--} & \approx & 0
\eea
These equations hold when inserted into any 2-point correlator,
and within the approximations of section \ref{approx}.

\sm

The equations $\cT_{+-} \approx 0$, $\cJ_- \approx 0$, together with the chiral
conservation equations imply the general structure of the remaining 2-point
correlators. To see how this works, we consider first the case $q_+=0$, and
use these equations to derive the general structure of (\ref{91a}). For example,
by Lorentz invariance, the 2-point correlator of $\cJ_+$ must be of the form,
\bea
\label{91ab}
\left \< \cJ_+ (p) \, \cJ_+ (-p) \right \> \Big |_{q=0}
& \sim &  {p_+ \over p_-} f(p^2)
\eea
where $f$ depends only on the Lorentz invariant $p^2=p_+p_-$. Following the
approximations of point 2. of section \ref{approx}, we neglect all fractional powers
of $p^2$, so that $f$ has a Laurent expansion in positive and negative
integer powers of $p^2$. Strictly positive powers produce local terms, which are to be
discarded following point 1. Strictly negative powers are prohibited by
the chiral conservation equation $p_- \cJ_+ (p)=0$ of (\ref{97b}), which
implies that $f$ must be constant. The value of the constant is derived
by explicit calculation \cite{D'Hoker:2010hr}.
The same argument may be adapted to the remaining correlators of (\ref{91a}), using the
chiral conservation of the stress tensor $p_- \cT_{++}(p)=p_+ \cT_{--}(p)=0$
within the approximations of section \ref{approx}.

\subsection{Structure of correlators for the charged magnetic brane}

We will now derive the general structure of the 2-point correlators of $\cJ$ and $\cT$,
constrained by extended charge conjugation and extended Lorentz symmetry, in the
approximation of section \ref{approx}, and under the (mild) assumption that
the $q_+ \to 0$ limit is smooth and leads to the general form (\ref{92b})
for all 2-point correlators.

\sm

The relations $\cT_{+-} \approx 0$ and $\cJ_- \approx 0$ lead to the vanishing of
all 2-point functions involving these operators, as was already expressed in
(\ref{91n}) and (\ref{93b}). The only remaining non-chiral correlator is
$\< \cT_{++} (p) \cT_{--} (-p) \>$; its general form is given by (\ref{92a}) and (\ref{92b}) with
$s_+=s_-=0$. The chiral conservation equations $p_+ \cT_{--} \approx p_- \cT_{++} \approx 0$
readily exclude all terms with $n \geq 1$, so that this correlator must vanish,
\bea
\< \cT_{++} (p) \cT_{--} (-p) \> \sim 0
\eea
The remaining correlators are all chiral.

\sm

For negative chirality, only a single 2-point correlator,  namely $\< \cT_{--} (p) \cT_{--} (-p) \>$,
remains in view of $\cJ_- \approx 0$ and (\ref{93b}).
Its general structure is given by (\ref{92a}) and (\ref{92b}) with $s_+ =0$ and $s_-=4$.
Chiral conservation $p_+ \cT_{--} \approx 0$ forces us to restrict to $n =1$.
On the other hand, all terms with $m >0$ are to be omitted by point 4. of the
approximations of section~\ref{approx}. Hence only the $m=0$, $n=1$ term survives,
which is independent of $q_+$ and thus must coincide with the result from the
purely magnetic case,
\bea
\left \< \cT_{--} (p) \, \cT_{--} (-p) \right \>
& \sim & {c \over 48 \pi} \, {p_-^3 \over p_+}
\eea

\sm

For positive chirality, we find the following 2-point correlators,
\bea
\label{92e}
\left \< \cJ_+ (p) \, \cJ_+ (-p) \right \>  & \sim & { kc \over 2 \pi} \, {p_+ \over p_-}
\no \\
\left \< \cJ_+ (p) \, \cT_{++} (-p) \right \>  & \sim & \g _1 {q_+ p_+ \over p_-}
\no \\
\left \< \cT_{++} (p) \, \cT_{++} (-p) \right \>
& \sim &
{c \over 48 \pi} \, {p_+^3 \over p_-}  +  \g_2 \, {q_+^2p_+ \over p_- }
\eea
where $\g_1$ and $\g_2$ are two coefficients which are not determined by
general arguments, and which will have to be obtained by explicit calculation.
The terms proportional to $c$ on the first and last lines correspond to the $q_+=0$
contribution from  (\ref{91a}).

\sm

To prove the form of the correlator $\< \cJ_+ \cJ_+\>$ in (\ref{92e}), we
set $s_+=2 $ and $s_-=0$ in (\ref{92a}), and use chiral conservation $p_- \cJ_+(p) \approx 0$
to restrict the expansion of  $F$ in (\ref{92b}) to $n  \leq 2 $ and $n \leq m+1$.
The $n=1$ contribution is local for $m \geq 1$, while the
$n=2$ contribution is always local.\footnote{Note that a contribution in $(p_-)^{-n}$ with $n \geq 1$ is
local, even though it is not polynomial in $p_\pm$.} Hence only the $n=1$, $m=0$ contribution
remains, which is independent of $q_+$ and coincides with the first line of (\ref{91a})
which yields the first line of (\ref{92a}).

\sm

The form of the correlator $\< \cJ_+ \cT_{++} \>$ in (\ref{92e}) is proven in analogous
fashion. Its general form is given by (\ref{92a}) and (\ref{92b}) with $s_+=3, s_-=0$.
Chiral conservation of either $\cJ_+$ or $\cT_{++}$ forces $n  \leq 3 $ and $n \leq m+1$.
Extended charge conjugation requires $m$ to be odd, so that we must actually have
$m \geq 1$ and $n \geq 2$. The contribution $n=3$ is local, leaving only $n=2$ with
$m=1$, with all $m\geq 3$ contributions to be omitted in view of point 4. of section~\ref{approx}.
This is precisely the form given in the second line of (\ref{92e}).

\sm

The form of the correlator $\< \cT_{++} \cT_{++} \>$ in (\ref{92e}) is given by (\ref{92a})
and (\ref{92b}) with $s_+=4$ and $ s_-=0$.
Chiral conservation of $\cT_{++}$ forces $n  \leq 4 $ and $n \leq m+1$.
Extended charge conjugation requires $m$ to be even, which restricts to
$m=0,2$. All contributions with $m-n \geq 1$ are to be omitted in view of point 4.
of section \ref{approx}, while the $n=4$ contributions, as well as the $n=m=2$
contributions are local. This leaves only the contribution $m=0, ~n=1$ corresponding to the
first term of the correlator in (\ref{92e}), and the contribution $m=2, ~ n=3$ corresponding to the second
term in (\ref{92e}). Since the first term is independent of $q_+$, its normalization is
provided by the result of (\ref{91a}).

\sm

Explicit evaluation in section \ref{seven} will confirm the general form of the
2-point correlators derived above, and will determine the values of the
coefficients as follows,
\bea
\label{92f}
\g _1 = -{  c_V c_E \over 4 \pi G_3 q_+ }
\hskip 1in
\g_2 = { c_V^2 c_E^2 \over 4 \pi G_3 kb (q_+)^2 }
\eea
The coefficients are independent of $q_+$ since $c_E/q_+$ is.
We will also show in section \ref{seven} that the 2-point correlators in (\ref{92e})
 remain the dominant IR contribution throughout  $k > 1/2$.

\subsection{Structure for general value of  $\hat B \geq \hat B_c$ and twisting}
\label{hatB}

The value of the normalized magnetic field $\hat B$, which was expressed in terms of the
parameters of the charged background solution in (\ref{23f}), is found to enter
the general structure of the 2-point correlator only through $\g_1$ and $\g_2$.
In particular, for the purely magnetic solution we have $\hat B = \infty$ and $q_+=0$,
so that  $\g_1=\g_2=0$, but for finite $\hat B$ we have $\g_1, \g_2 \not= 0$.
Thus, the two Virasoro algebras and the chiral
U(1)-current algebra, which were identified for the purely magnetic case, persist
for all $\hat B \geq \hat B_c$.
As the charge density is turned on, however, the current
and stress tensor operators become {\sl twisted}. In the positive chirality sector,
this may be seen by diagonalizing the system of 2-point functions of $\cJ_+$ and $\cT_{++}$,
by setting,
\bea
\label{94a}
\cJ _+ & = & \cJ_+ ^{(0)}
\no \\
\cT_{++} & = & \cT_{++} ^{(0)} - \mu_+ \cJ_+ ^{(0)}
\hskip 1in \mu_+ = { c_V c_E \over kb} = - {2 \pi \over kc} \g_1 q_+
\eea
If $\cJ_+^{(0)}$ and $\cT_{++}^{(0)}$ obey the 2-point functions of the
purely magnetic solution of (\ref{91a}), then $\cJ_+$ and $\cT_{++}$
defined by (\ref{94a}) will obey (\ref{92e}). Note that this result
crucially depends upon the precise value taken by $\g_2$, and its
relation to $\g_1$, given by,
\bea
\g_2  = { 2 \pi \over kc} \, \g_1^2
\eea
Note that the twisting of (\ref{94a}) is possible because both $\cJ_+^{(0)}$ and $\cT_{++}^{(0)}$
are chirally conserved.

\newpage

\section{Linearized metric and gauge field fluctuations}
\setcounter{equation}{0}
\label{five}

In this section, we begin the process of calculating the two-point correlators
of the $U(1)$ current $\cJ$ and the stress tensor $\cT$. To do so, the full
Einstein-Maxwell-Chern-Simons equations are linearized around the charged
magnetic background solution. Identifying the source and vev
components in the Fefferman-Graham expansion of these solutions,
we derive the vevs of the current and stress tensor using (\ref{20d}),
and from there extract (the non-local parts of) all two-point functions.
Since the background solution is invariant
under translations in $x^\mu$, we solve for plane waves with fixed momenta
$p_\mu$ for $\mu = +, -, 1,2$. For simplicity, we restrict to the most
interesting case where $p_1=p_2=0$, leaving only the momenta $p_\pm$.

\subsection{Parametrization of linear fluctuations}

A general Ansatz for the plane wave fluctuations of the metric and gauge field
is given by,
\bea
\label{40a}
A & = & A_C  + a_M \, e^{ipx} dx^M
\no \\
ds^2 & = & ds_C^2  + h_{MN} \, e^{ipx} dx^M dx^N
\eea
We continue to use the notation $px=p_+ x^+ + p_- x^-$.
Here, $A_C$ and $ds_C^2 $ are the gauge potential and metric of the
charged magnetic brane solution. In the gauge adopted in
(\ref{21a}),  referred to as \ban gauge,
the charged background solution takes the form,
\bea
\label{40aa}
ds_C^2 & = & {dr^2 \over L_0(r)^2} + M_0(r) (dx^+)^2 + 2 L_0(r) dx^+ dx^-
+ e^{2V_0(r)} dx^i dx^i
\no \\
A_C & = & b x^1 dx^2 + A_0(r) dx^+
\eea
where the functions $L_0$ and $V_0$ are those of the purely magnetic background solution,
obeying equations E1, E4, and fV of  (\ref{21aa}) with $N=P=0$, while $A_0$ and $M_0$
are given in (\ref{21d}).
Translation invariance in $x^\mu$ of the solution (\ref{40aa}) justifies our
consideration of plane wave fluctuations in (\ref{40a}). The functions $a_M$ and
$g_{MN}$ depend on the holographic radius, but are independent of $x^\mu$.
Invariance under rotations in the $x^{1,2}$ plane of the solution (\ref{40aa})
leads us to require rotation invariance of the fluctuations of (\ref{40a}),
which implies the following conditions,
\bea
\label{40b}
a_i = h_{i r}=h_{i+}=h_{i-}=0 \hskip 0.8in h_{ij} = \delta _{ij} h_K
\hskip 0.5in i,j=1,2
\eea
Even with the benefit of translation and rotation symmetry, the corresponding
reduced field equations remain quite involved, and will not be presented here.

\sm

To obtain workable reduced equations we choose a gauge.
It turns out that a convenient gauge choice is not global, but is rather obtained
by patching together the light-cone gauge for the near region and the
\ban gauge for the far region. In the overlap region, a gauge transformation
is required to patch the two gauge choices together. In the extreme UV limit of the
far region, we will make a further gauge transformation to Fefferman-Graham
gauge, which is required to properly extract the current and stress tensor data.

\subsubsection{Near-region: light-cone gauge}

In the near region, defined by $r\ll1$, we will adopt light-cone gauge,
and denote the corresponding fluctuation fields with tildes, namely
$\tilde a_M$ and $\tilde h_{MN}$. These fields satisfy the usual light-cone
gauge conditions,
\bea
\label{40c}
\ti a_- = \ti h_{r-} = \ti h_{+-} = \ti h_{--}=0
\eea
in addition to the consequences of invariance under  translations in $x^\mu$,
and rotations in $x^1,x^2$ as in (\ref{40b}). The full gauge field and metric in light-cone gauge become,
\bea
\label{40d}
\ti A & = & A_C  + \Big ( \ti a_r  dr + \ti a_+  dx^+ \Big ) \, e^{ipx}
\no \\
d\ti s^2 & = & ds_C^2  + \Big (
\ti h_{rr} dr^2 + 2 \ti h_{r+} dr dx^+ + \ti h_{++} (dx^+)^2 + \ti h_K dx^i dx^i \Big ) \, e^{ipx}
\eea
where the remaining fields $\ti a_r, \ti a_+, \ti h_{rr}, \ti h_{r+}, \ti h_{++}$, and
$\ti h_K$ depend only on $r$.

\subsubsection{Far region: \ban gauge}

In the far region all momentum-dependence in the reduced field equations
is to be neglected. This specification always requires $p_+p_- \ll r$.
As we have discussed for the case of scalar fluctuations in section 3.2, however,
depending on the value of $k$, additional restrictions on $r$ may be needed
in order to ensure that the far region will extend non-trivially into the near region.
We will exhibit these conditions as we proceed forward. We stress, however,
that even if the reduced field equations do not involve momenta in the far region, their
boundary conditions will be given by $p$-dependent coefficients.

\sm

When all momentum dependence in the reduced equations is neglected,
the fluctuations will be analyzed in \ban gauge, where they correspond to
$x^\mu$-independent fluctuations of the functions $L,M,N,V,A_\pm$
of (\ref{21a}). \ban gauge will be convenient here, because this linear
fluctuation problem was considered and solved already
in Section 5 and Appendix B of \cite{D'Hoker:2010ij}.
The fluctuation fields in \ban gauge will be denoted by the original letters
$a_M$ and $h_{MN}$, and are subject to  the following  gauge conditions,
\bea
\label{4e}
a_r=h_{r+} = h_{r-} =0
\hskip 0.8in
h_{rr} = - {2 \over L_0^3} h_{+-}  + {M_0\over L_0^4} h_{--}
\eea
in addition to the consequences of invariance under translations in $x^\mu$,
and rotation in $x^1,x^2$ as in (\ref{40b}). The relation for $h_{rr}$ is imposed to guarantee that
the form of the $dr^2$ term is preserved by the fluctuations.
Thus, the gauge field and metric in \ban gauge become,
\bea
\label{40f}
A & = & A_C  + \Big ( a_+  dx^+ + a_-  dx^- \Big ) \, e^{ipx}
\\
ds^2 & = & ds_C^2  +
\Big ( h_{rr} dr^2 +  h_{++} (dx^+)^2 + 2 h_{+-} dx^+ dx^-
+  h_{--} (dx^-)^2 +  h_K dx^i dx^i \Big ) e^{ipx}
\no
\eea
The remaining functions $a_\pm, h_{rr}, h_{\pm \pm}, h_{+-}$, and $h_K$
depend on $r$, but are independent of $x^\mu$.

\subsection{Reduced field equations in the near region}

In the near region, where $r \ll 1$, we adopt light-cone gauge. The reduced
equations for fluctuations in light-cone gauge around the full charged
magnetic background solution are derived in Appendix B.
For $r\ll1$, it is appropriate to use the
approximation,\footnote{In keeping with the discussion of section \ref{twoseven},
we restrict attention to the critical background solution for which $\hat B = \hat B_c$,
and $\a_0=0$, so that no linear term
in $M_0$ is present as $r \to 0$. The solutions for $\hat B > \hat B_c$ may then be reconstructed by applying an $SL(2,R)$ transformation to the critical  solution.}
\bea
\label{41a}
V _0 (r) = 0 ~~ & \hskip 1in & A_0 (r) =  { q  r^k \over k}
\no \\
L_0 (r) = 2 br & \hskip 1in & M_0 (r) =  - { 2 q^2 r^{2k} \over k (2k-1)}
\eea
so that the general light-cone equations (\ref{B1}), (\ref{B2}),
(\ref{B3}), (\ref{B4}), and especially (\ref{B5}) greatly simplify.
The reduced equations may be presented most symmetrical  in terms of the
functions $\hti_K$ and $\cti (r) =  4 r \ati (r)$, and we have,
\bea
\label{41b}
&& 0 = r^2 \hti_K'' + 2r \hti_K' -{ 4 \over 3} \hti_K - { p^2 \over 4 b^3r} \hti_K
-{q^2 p_-^2 r^{2k-2} \over 72 k(2k-1) }\hti_K + { i qp_-  r^{k-1} \over b^3}   \cti_r
\no \\
&& 0= r^2 \cti_r'' + 2 r \cti_r' -k(k-1) \cti_r -{ p^2 \over 4b^3 r} \cti_r
-{ q^2 p_-^2 r^{2k-2} \over 72 k (2k-1)} \cti_r - {i  qp_- r^{k-1} \over b^3} \hti_K \qquad
\eea
Note that the equations only depend on the combinations $p^2=p_+p_-$ and $qp_-$.
For later use, we also record the relations giving the remaining fields in the
near region,
\bea
\label{41c}
\hti_{rr}  = - {\hti_K \over 6 r^2} \hskip 0.25in
& \hskip 0.7in &
\hti_{++}  = {288 r^2 \over p_-^2}
\left ( r^2 \hti_K'' + 2 r \hti_K' -{4 \over 3} \hti_K + { i \over b^3}  q p_- r^k \ati_r \right )
\no \\
\hti_{r+}  = -{ 4 i b r \over p_-} \hti_K'
& \hskip 0.7in &
\ati_+  = {24 i b r^2 \over p_-} \left ( (r \ati_r  )' +  k  \ati_r \right )
\eea
Rescaling $r$ by $p^2$,
\bea
\label{41d}
r={ p^2x \over b^3}  \hskip 1in \hti_K(r) = \hat h_K(x) \hskip 1in \cti_r (r) = \hat c _r (x)
\eea
reveals that the equations in terms of the rescaled variable $x$,
\bea
\label{41e}
0 &= & x^2 \hat h_K'' + 2x \hat h_K' -{ 4 \over 3} \hat h_K - { 1 \over 4 x} \hat h_K
-{1 \over 4} \xi^2 x^{2k-2}  \hat h_K +  i \xi_1 \xi  \, x^{k-1}  \hat  c_r
\no \\
0 & = & x^2 \hat c_r'' + 2 x \hat c_r' -k(k-1) \hat c_r -{1 \over 4x} \hat c_r
-{ 1 \over 4} \xi^2 x^{2k-2} \hat c_r -  i \xi_1 \xi \, x^{k-1}  \hat h_K
\eea
intrinsically only depend on the combination $\xi$, which was already
introduced in (\ref{35f}) for the scalar fluctuation problem.
Here, we have used the notation  $ \xi_1^2 = 2k(2k-1)/3$.

\sm

On the one hand, for $k > 1$, we automatically have $ \xi \to 0$ in the limit of
small momenta $p_\pm$, no matter how this limit is being taken. Also,
the $\xi$-dependent terms in (\ref{41e}) are well-behaved as $x \to 0$.
On the other hand, for $1/2 < k < 1$, the combination $\xi$ exhibits a
directional singularity in the limit of small momenta $p_\pm$, so that
$\xi$ may tend to zero in one limit, to $\infty$ in another, or to
any finite value. Also, for $1/2<k<1$, the $\xi$-dependent
terms in (\ref{41e}) diverge as $x \to 0$. Finally, it is unlikely that
equations (\ref{41e}) can be solved for finite $\xi$, beyond the
type of perturbative results that were derived for the scalar problem
in Appendix A. To avoid these three
complications, we will henceforth restrict to the case where $k \geq 1$,
and proceed to solve it in the limit of low momenta, which implies $ \xi \ll 1$.

\subsection{Solutions in the near region}

Assuming $ k \geq 1$ and $\xi \ll 1$, equations (\ref{41b}) decouple from one another,
\bea
\label{42a}
0&= & r^2 \hti_K'' + 2r \hti_K' -{ 4 \over 3} \hti_K - { p^2 \over 4r} \hti_K
\no \\
0 & = & r^2 \cti_r'' + 2 r \cti_r' -k(k-1)\cti_r -{ p^2 \over 4 r} \cti_r
\eea
and may be solved in terms of modified Bessel functions,
\bea
\label{42b}
\hti_K (r) & = & { v_0 \over \sqrt{r}}
\left ( I_{2 \sigma +1} \left ( \sqrt{{ p^2 \over b^3 r}} \right )
- I_{-2 \sigma -1} \left ( \sqrt{{ p^2 \over b^3 r}} \right ) \right )
\no \\
\cti_r (r) & = &  { c_0 \over \sqrt{r}}
\left ( I_{2 k-1} \left (\sqrt{{ p^2 \over b^3 r}} \right )
- I_{-2k +1} \left ( \sqrt{{ p^2 \over b^3 r}} \right ) \right )
\eea
where  $v_0$ and $c_0$ are independent of $r$, and $\sigma $ was
given in (\ref{23c}). As was the case for the scalar fluctuations, the
above linear combinations of Bessel functions are required
by regularity of $\hti$ and $\ati$ as $ r \to 0$, and are proportional to
$K_{2 \sigma +1}$ and $K_{2k-1}$ respectively.

\subsection{Solutions in the overlap region (light-cone gauge)}

The overlap region, defined by $p^2 \ll r \ll 1$,  is contained in the
near region so that the approximations of (\ref{41a}), as well as
the reduced equations of (\ref{42a}) are valid. It is also contained
in the far region $p^2 \ll r$, where all momentum dependence in the
reduced field equations is to be neglected, so that equations (\ref{42a})
actually reduce to,
\bea
\label{41b}
0 & = & r^2 \hti_K'' + 2r \hti_K' -{ 4 \over 3} \hti_K
\no \\
0 & = & r^2 \cti_r'' + 2 r \cti_r' -k(k-1) \cti_r
\eea
Its solutions are given by,
\bea
\label{43a}
\hti_K(r) & = & v_+ r^\sigma + v_- r^{-1-\sigma}
\no \\
\ati_r(r) & = & {1 \over 4} \left ( c_+ r^{k-2} + c_- r^{-k-1} \right )
\eea
The coefficients $v_\pm$ and $c_\pm$ are independent of $r$, and
may be determined in terms of $v_0$ and $c_0$ respectively by
matching the solutions of (\ref{43a}) with those of (\ref{42b}) in the
$p^2 \ll r$ limit. Here, we will only need the ratios of these coefficients,
\bea
\label{43c}
{v_- \over v_+} = - {\G (- 2 \sigma) \over \G (2+2 \sigma)}
\left ( { p\over 2} \right )^{4 \sigma +2}
\hskip 1in
{c_- \over c_+} = - {\G (2- 2k) \over \G (2k)}
\left ( { p\over 2} \right )^{4k -2}
\eea
Clearly, to leading order in small momenta, the coefficients
$v_- $ and $c_-$ are suppressed compared to $v_+$ and $c_+$,
and the corresponding terms may effectively be dropped. In terms of the
coefficients $v_\pm$, $c_\pm$, the remaining light-cone gauge functions
are given as follows,
\bea
\label{43d}
\hti_{rr} = - {\hti _K(r) \over 6 r^2}
& \hskip 1in &
\hti_{r+} = - {4i b \over p_-} \left ( \sigma v_+ r^\sigma - (\sigma +1) v_- r^{-\sigma -1} \right )
\no \\
\hti_{++} = 0 \hskip 0.5in
&&
\ati_+ =  { 6 i b (2k-1) \over p_-}  c_+ r^k
\eea
The cancellation of $\hti_{++}$ is brought about by the fact that its various terms
in (\ref{41c}) cancel for the overlap region solution of $\hti _K$ of (\ref{43a})
and/or are of higher order in $p^2/r$.

\subsection{Change to \ban gauge in overlap region}

We have solved the linearized equations around the charged magnetic brane
in the near region in light-cone gauge, and need to match the result with the solution in the
far region in \ban gauge. The matching takes place in the overlap region
$p^2 \ll r \ll 1$ where both approximations are simultaneously valid, and where
a simplified solution holds. The general form of the solution in light-cone gauge
will be denoted by,
\bea
\label{60a}
\ti A_M dx^M & = & A_C  + \ti a_M (r) \, e^{ipx} dx^M
\no \\
\ti H_{MN}dx^M dx^N & = & ds_C^2  + \ti h_{MN} (r) \, e^{ipx} dx^M dx^N
\eea
while its general form in \ban gauge will be denoted by,
\bea
\label{60b}
A_M dx^M & = & A_C + a_M (r) \, e^{ipx} dx^M
\no \\
H_{MN}dx^M dx^N & = & ds_C^2  + h_{MN} (r) \, e^{ipx} dx^M dx^N
\eea
Since the background solution is the same in both
gauges, the gauge transformation between the linear fluctuations may be
carried out at the linearized level. This involves a diffeomorphism vector
field $U^M$ and a gauge transformation field $\Theta$ which take the form,
\bea
\label{60c}
U^M (r,x) & = & u^M (r) \, e^{ipx}
\no \\
\Theta (r,x) & = & \theta (r) \, e^{ipx}
\eea
To convert the metric and gauge fluctuations $\Hti_{MN}$ and $\Ati_M$
of the light-cone gauge to $H_{MN}$ and $A _M$ of \ban
gauge in the overlap region, we solve the equations,
\bea
\label{60d}
H_{MN} & = & \Hti_{MN} + \nabla _M U_N + \nabla _N U_M
\no \\
A_M & = & \Ati_M + U^K F_{KM} + \p_M \Theta
\eea
for $U^M$, $\Theta$, as well as $H_{MN} $ and $A_M$.
The covariant derivatives $\nabla _M$, and the field strength $F_{KM}$
are with respect to the
background solution, and so is the metric used to raise and lower indices.
Substituting the plane wave forms of (\ref{60a}) and (\ref{60b}),
\bea
\label{60e}
h_{MN} e^{ipx} & = & \hti_{MN} e^{ipx} + \nabla _M U_N + \nabla _N U_M
\no \\
a_M e^{ipx} & = & \ati_M e^{ipx} + U^K F_{KM} + \p_M \Theta
\eea
Imposing rotation invariance, as spelled out in (\ref{40b}), on $\hti_{MN}$, $\ati_M$,
$h_{MN}$, and $a_M$ gives $u^1=u^2=0$. Next, we impose the conditions for
light-cone gauge of (\ref{40c}) on $\hti_{MN}$ and $ \ati_M$, and
for \ban gauge of (\ref{4e}) on $h_{MN}$, and $a_M$. Finally, we assign the
expressions for the light-cone gauge solution in the overlap region of (\ref{43a})
and (\ref{43d}) to the remaining components of the fields $\hti_{MN}$ and $\ati_M$.
The resulting equations are derived in (\ref{C2}), (\ref{C3}), and (\ref{C4})
of Appendix \ref{appC}.

\subsection{Solutions in the overlap region (\ban gauge)}

The corresponding solutions in the overlap region  for the metric and
gauge fluctuations in \ban gauge are constructed
in Appendix \ref{appC} as well, and are of the following form,
\bea
\label{61a}
h _{++}(r) & = & s_{++} r + t_{++}
\no \\
h _{--} (r) & = & s_{--} r + t_{--}
\no \\
h_{+-} (r) & = & s_{+-} r + t_{+-}
\no \\
a_+ (r) & = &  \s_+  + \tau_+ r^{k}
\no \\
a _- (r) & = & \s_- + \tau_- r^{-k}
\eea
with the coefficients given by,
\bea
\label{61b}
s_{++} = 4 i b p_+ u_0^- & \hskip .8in &
t_{++}  =  { p_+ p_- \over 24 b} s_{++} +{ p_+^3 \over 24 b p_-} s_{--}
\no \\
s_{--} = 4 i b p_- u_0^+  &&
t_{--}  =  { p_+ p_- \over 24 b} s_{--} +{ p_-^3 \over 24 b p_+} s_{++}
\no \\
s_{+-}=0 \hskip 0.5in &&
t_{+-} = 2 b u^r_0 +{1 \over 24 b} \left ( p_+^2 s_{--} + p_-^2 s_{++} \right )
\no \\
\s_+ = ip_+ \theta _0 \hskip 0.21in & \hskip .8in &
\tau _+ =  (2k-1) {6ib c_+ \over p_-}  + {qp_+ (k-1) \over 4b p_- k}  s_{--}
+ { q p_- \over 4  b p_+} s_{++}
\no \\
\s_- =  i p_- \theta _0 \hskip 0.21in && \tau _- =  {i p_- c_-  \over 4k}
\eea
Since $c_-/c_+ \sim p^{4k}$, we may effectively set $c_-=0$ within the
approximations of section \ref{approx}. The metric fluctuations of the field $h_K$ are given by,
\bea
\label{61c}
h_K(r) =  v_+ r^\sigma + v_- r^{-1-\sigma}
\eea
and decouple from all other fluctuations of both the metric and the gauge field.
Since $v_-/v_+ \sim p^{4 \sigma +2}$, we may effectively set $v_-=0$
within the approximations of section \ref{approx}.

\subsection{Reduced field equations in the far region}

In the far region, defined by $p^2 \ll r$, all momentum dependence in the field
equations is to be omitted. The corresponding reduced field equations
for the metric and gauge field fluctuations, in \ban gauge, coincide with
the perturbative expansion of the fields $L,M,N,V,E,P$, obtained already in
equations (5.35), (5.36) and (5.37) of \cite{D'Hoker:2010ij}.
Making the correspondence with the present notations, the expansion
is as follows,
\bea
\label{62a}
L(r,x) & = & L_0(r) + \ep L_1(r) \, e^{ipx} \hskip 1.05in L_1 (r)= h_{+-} (r)
\no \\
M(r,x) & = & M_0(r) + \ep M_1(r) \, e^{ipx} \hskip 0.92in M_1 (r) =h_{++} (r)
\no \\
N(r,x) & = &  \ep N_1(r) \, e^{ipx} \hskip 1.6in N_1(r)  = h _{--}(r)
\no \\
E(r,x) & = & E_0(r) + \ep E_1(r) \, e^{ipx} \hskip 1.02in E_1 (r) = a_+(r)'
\no \\
P(r,x) & = &  \ep P_1(r) \, e^{ipx} \hskip 1.64in P_1(r)= - a_-(r)'
\no \\
V(r,x) & = & V_0(r) + \ep V_1(r) \, e^{ipx} \hskip 1.1in 2 e^{2V_0(r)} V_1 (r) = h_K (r)
\no \\
f(r,x) & = & f_0 (r) + \ep f_1 (r) \, e^{ipx}
\eea
where $\ep$ is a formal expansion parameter. In \cite{D'Hoker:2010ij}, we had
also introduced the gauge potentials $A_1$ and $C_1$, which are related
to the fields $E_1$ and $P_1$ by $E_1=A_1'$ and $P_1=C_1'$; these potentials
are related to the present fields by
\bea
\label{62b}
A_1 (r) & = & a_+(r) - A_+^0
\no \\
-C_1(r) & = &  a_-(r) - A_-^0
\eea
where $A_\pm ^0$ are constants. Here, $f$ was defined in (\ref{21bb}), and we have,
\bea
\label{62c}
f_0 & = & L_0^2
\no \\
f_1 & = & 2 L_0 L_1 - M_0 N_1
\eea
Maxwell's equations for the perturbation functions are given by,
\bea
\label{62d}
{\rm M1} & \hskip 0.4in &
0 = \left ( e^{2V_0} ( E_0 N_1 + L_0 P_1) \right ) ' + 2 k b P_1
\no \\
{\rm M2} &&
0 = \left ( e^{2V_0} ( 2L_0 E_0 V_1 + L_0 E_1 +E_0 L_1 +M_0 P_1) \right )' - 2 kb E_1
\eea
while Einstein's equations are given by,
\bea
\label{62e}
{\rm E1} & \hskip 0.3in &
0 = L_1'' + 2 V_0 ' L_1' + 2 V_1' L_0' + 4 L_1 \left ( V_0'' +(V_0')^2 \right )
+ 4 L_0 \left ( V_1'' + 2 V_0' V_1'  \right ) - 4 E_0 P_1
\no \\
{\rm E2 }  && 0 = M_1'' + 2 V_0 ' M_1' + 2 V_1' M_0' + 4 M_1 \left ( V_0'' +(V_0')^2 \right )
+ 4 M_0 \Big ( V_1'' + 2 V_0' V_1' \Big ) + 8 E_0 E_1
\no \\
{\rm E3} && 0 = N_1'' + 2 V_0 ' N_1' + 4 N_1 \left ( V_0'' +(V_0')^2 \right )
\no \\
{\rm fV} && 0 = f_1'' + 4 V_0' f_1' + 4 V_1' f_0' + 2 V_1'' f_0
+ 2 V_0'' f_1 + 4 (V_0')^2 f_1 + 8 V_0' V_1' f_0
\no \\
{\rm E4} && 0 = \left (6V_0'' + 12 (V_0')^2 \right ) f_1
+ \Big (6 V_1'' + 24 V_0 ' V_1' \Big ) f_0  + 6 V_1' f_0' + 6 V_0' f_1'
\no \\ && \qquad
 - 32b^2 e^{-4V_0} V_1 + 8 L_0 E_0 P_1 + 4 N_1 E_0^2
\eea
We also have first integral equations, of which we will need
the following,
\bea
\label{62f}
e^{2V_0} ( E_0 N_1 + L_0 P_1)  & = & - 2 k b C_1
\no \\
e^{2V_0} ( 2L_0 E_0 V_1 + L_0 E_1 +E_0 L_1 +M_0 P_1)  & = & 2 kb A_1
\no \\
(N_1M_0'-M_0N_1')e^{2V_0} - 8kb A_0C_1 & = & 2 \lambda _0
\no \\
(N_1L_0'-L_0N_1')e^{2V_0}  & = &  \nu _0
\eea
The solution to these equations in the far region needs to be matched
to the solution of (\ref{61a}), (\ref{61b}), and (\ref{61c}) that we have
derived earlier for the overlap region (recall that the overlap region
is contained in the far region, so that both solutions must match
in the overlap region). Thus, in the overlap region, we must have,
\bea
\label{62g}
L_1(r) & = & s_{+-} r + t_{+-} \hskip 1in s_{+-}=0
\no \\
M_1 (r) & = & s_{++} r + t_{++}
\no \\
N_1(r) & = & s_{--} r + t_{--}
\no \\
A_1(r) & = & \s_+ + \tau_+ r^k
\no \\
-C_1(r) & = & \s_-  + \tau_- r^{-k}
\no \\
2e^{2V_0(r)} V_1(r) & = & v_+ r^\sigma + v_- r^{-\sigma -1}
\eea
where the coefficients $s_{\pm \pm}, t_{\pm \pm}, s_{+-}, t_{+-}, \s _\pm$,
and $\tau _\pm$ are related by the equations of (\ref{61b}). The coefficients
$v_\pm$ are  decoupled from all other fields. At low momenta, the coefficient $v_-$
may effectively be set to 0.

\subsection{Solutions in the far region}

The system of linear equations of (\ref{62d}) and (\ref{62e}) for the far region
turns out to be solvable by quadratures in the following sense. Their general
solution may be obtained by (successive) quadratures in terms of the
functions $L_0$ and $V_0$ which characterize the purely magnetic solution.
In fact, $L_0$ itself may be obtained in terms of $V_0$ by quadratures.
Thus, although $V_0$ is not (as of yet) known analytically, all other
fluctuation functions are determined explicitly in terms of $V_0$.
The system may be integrated iteratively in the following sequence,
\bea
\label{63aa}
{\rm E3} ~ \longrightarrow ~ {\rm M1} ~ \longrightarrow & ({\rm fV}, {\rm E4}) & \longrightarrow ~ {\rm M2} ~ \longrightarrow ~ {\rm E2}
\no \\
N_1 ~ \longrightarrow ~~ P_1 ~ \longrightarrow & (f_1,V_1) & \longrightarrow ~ E_1 ~~
\longrightarrow ~ M_1
\eea
The top line gives the order in which the equations should be solved, and the
bottom line gives the corresponding functions obtained. Equations fV and
E4 are coupled and must be solved together; their solution is obtained
by solving a single auxiliary third order linear differential to which two solutions
are a priori known for symmetry reasons. The full solution in the far region
was obtained in \cite{D'Hoker:2010ij}.

\sm

In preparation for the calculation of correlators, we will review  here
the solution for the functions $N_1$ and $P_1$,
which do not require the more complicated solution for $V_1$ and $f_1$.
These solutions were obtained already in \cite{D'Hoker:2010ij}. Inspection of  equation E3
for $N_1$ in (\ref{62e}) and equation E1 of (\ref{21aa}) shows that
one solution for $ N_1$ is proportional to $L_0$, while the other (linearly
independent) solution is proportional to the function $L_0^c$, defined by,
\bea
\label{63a}
L_0 ^c (r) = L_0(r) \int ^r _ \infty { dr' \over L_0(r')^2 e^{2V_0(r)}}
\eea
The function $L_0^c$ involves data of the purely magnetic background solution
only, and has the following asymptotics,
\bea
\label{63b}
r \to 0 ~  & \hskip 0.8in & L_0^c(r) = -{1 \over 2b} + \cO(r^\sigma)
\no \\
r \to \infty && L_0 ^c(r) = - {1 \over 4 c_V (r-r_0)} + { 3 \ln r \over 16 c_V^3 \, r^3}
+ \cO(r^{-3})
\eea
Thus, the general solution for $N_1$ is a linear combination of $L_0$
and $L_0^c$. Since we have $L_0(r) = 2 br$ for small $r$, we see that
the asymptotic behavior of $N_1(r)$  is indeed of the form given already
in (\ref{62g}) for the overlap region. Identifying the coefficients gives,
\bea
\label{63c}
N_1(r) = { L_0(r) \over 2b} s_{--}  - 2 b L_0^c(r) t_{--}
\eea
Since this solution does not involve the charge density parameter $q$, it
coincides with the corresponding fluctuation solution for the purely magnetic
case of \cite{D'Hoker:2010hr}.

\sm

Next, the function $P_1$, and thus $A_-$, may be obtained by solving equation M1.
It is actually more convenient to solve instead the first integral equation on the
third line of (\ref{62f}), since this equation gives directly $C_1=A_- ^0 -A_-$ in terms of
data that we have already obtained. The result is as follows,
\bea
\label{63d}
A_-(r) = \sigma _-  + {\lambda _0 \over 4 kb A_0(r)}
+ { A_0(r) \over 4b}s_{--} - b  {A_0L_0^c \over L_0}(r) t_{--}
-   {M_0 \over 4k L_0 A_0}t_{--}
\eea
Using $A_0(r) \sim qr^k/k$ as $r \to 0$, we match the asymptotics of this
expression with the one in the overlap region given in (\ref{61a}).
This accounts for the presence of the additive constant $\sigma _-$ in (\ref{63d}),
and solves for  $\lambda _0=4 b q \tau_-$. Since the coefficient
$\tau_- \sim c_- \sim p^{4k}$ is suppressed within our approximation, we will set
\bea
\label{63f}
\tau_- = \lambda _0=0
\eea
in (\ref{63d}), and throughout the remainder of this paper.

\newpage

\section{Current and stress tensor correlators: general set-up}
\setcounter{equation}{0}
\label{six}

In this section, we will begin the calculation of the actual 2-point correlators of the
current~$\cJ$ and the stress tensor~$\cT$ in the state dual to the charged
magnetic solution. A number of general implications will be derived
from the structure of the sources and the solutions in the far and overlap regions,
which will form the basis for the computation of all correlators in later sections.

\subsection{Identification of sources and vevs}
\label{sixone}

Sources and vevs may be identified from the $r \to \infty$ boundary asymptotics
of the metric and gauge fluctuations, following (\ref{20b}) and (\ref{20c}),
\bea
\label{52a}
L_1 (r) & \sim & 4 r \go _{+-} + g^{(2)}_{+-} + {1 \over 4r} g^{(4)} _{+-} + \cdots
\no \\
M_1 (r) & \sim & 4 r \go _{++} + g^{(2)}_{++} + {1 \over 4r} g^{(4)} _{++} + \cdots
\no \\
N_1 (r) & \sim & 4 r \go _{--} + g^{(2)}_{--} + {1 \over 4r} g^{(4)} _{--} + \cdots
\no \\
A_1 (r) & \sim &  \Ao _+ + {1 \over 4 r} \At _+ + \cdots
\no \\
-C_1 (r) & \sim &  \Ao _- + {1 \over 4 r} \At_- + \cdots
\no \\
2 V_1 e^{2V_0(r)} & \sim & 4 r \go _{ii} + g^{(2)}_{ii} + {1 \over 4r} g^{(4)} _{ii} + \cdots \hskip 0.5in i=1,2
\eea
The factors of 4 in the normalization of $r$
have been included to ensure that the conformal boundary
metric has the proper customary normalization, given by,
\bea
\label{52b}
ds_0^2 = dx^+ dx^- + \beta (dx^+)^2 + {c_V \over 4} dx^i dx^i
\hskip 1in
\beta = - {\a \over b}
\eea
which requires the following relation between $r$ and the Fefferman-Graham
coordinate $\rho$,
\bea
\label{52c}
\rho = 4 (r-r_0) + { \ell _1 \over r} -{ 3+ 6 \ln r \over c_V^2 r} + \cdots
\eea
where $r_0$ and $\ell_1$ are constants which will not be needed here.
The  current and stress tensor vevs are given in terms
of $g^{(4)}_{\mu \nu} $ and $A^{(2)}_\mu $ by equation (\ref{20d}).
As in \cite{D'Hoker:2010hr}, we will prefer to deal with normalized 2-dimensional
current and stress tensor, obtained by integrating over the $x^{1,2}$
space with coordinate volume $V_2$, and effective 3-dimensional
Newton constant $G_3=G_5/V_2$. The expressions for the vevs of these
{\sl normalized current and stress tensor} are  given by,
\bea
\label{52cc}
16 \pi G_3  T_{\mu\nu}(x) & = & c_V \gf_{\mu\nu}(x) +{\rm local}
\no \\
8 \pi G_3  J_\mu(x) & = & c_V \At_\mu(x)+{\rm local}
\eea

\sm

We close this subsection by making a key observation. {\sl The sources
to the metric and gauge field fluctuations, $\go_{\mu \nu}$ and $\Ao_\mu$
are independent of the background charge density $q_+$.} Indeed, the
sources are external fields whose strength may be dialed from the outside
without reference to the value of $q_+$. This important
observation will be used throughout.

\subsection{Fluctuations in $V_1$ and correlators with $\cT_K$}
\label{six2}

In this section, we will derive the 2-point correlators involving the stress tensor
component $\cT_K = \cT_{11}=\cT_{22}$ conjugate to the metric variations
in $V_1$. Using the general result that the trace of the full stress tensor $\cT$,
defined in (\ref{91ca}) is governed by the trace anomaly, and has local correlators with
all operators, we will derive the result of (\ref{91cc}) for the two-dimensional trace
part $\cT_{+-}- \beta \cT_{--} \approx 0$.

\sm

The solutions in which $\cT_K$ is sourced are governed by the fluctuation
equations for $f_1, V_1$. These equations were solved in Appendix B of
\cite{D'Hoker:2010ij}, and the general solution takes the form,
\bea
V_1 = \zeta _t V_1^t + \zeta _d V_1 ^d + \zeta _n V_1 ^n +  V_1^p
\eea
where the scripts $t,d,n$ refer to the three homogeneous solutions for which $N_1=P_1=0$,
while $p$ stands for the inhomogeneous solution arising from $N_1 , P_1 \not= 0$.
Without loss of generality, the latter may be normalized so that $V_1^p$ vanishes
at $r=\infty$ and $r \to 0$. The mode $V_1^n (r)$ behaves as $r^{-1-\sigma}$
for $r \to 0$, matches onto the coefficient $v_-$ in the overlap region, and is thus
to be omitted within our approximation of section \ref{approx}, so that $\zeta _n=0$.

\sm

In summary, the stress tensor $\cT_K$ is sourced only by the {\sl dilation mode} $V_1^d (r)$,
so that we set $\zeta _t=\zeta _n=N_1=P_1=0$. The remaining
fluctuations for the dilation mode are given by,
\bea
L_1 ^d (r) & = & r L_0 (r)' - L_0(r)
\no \\
M_1^d (r) & = & r M_0(r)' - M_0(r)
\no \\
E_1 ^d (r) & = & r E_0(r)' - E_0 (r)/2
\no \\
V_1 ^d (r) & = & r V_0(r)'
\eea
This is an exact solution to all orders in $q_+$. Since $N_1(r)=0$, and
$M_1 ^d (r) \sim r^{2k}$ as $r \to 0$, the metric matching conditions
in the overlap region imply that we must have,
\bea
s_{\pm \pm } = t_{\pm \pm } =0
\eea
Since $L_1 ^d(r) \sim r^{1+\sigma}$, $E_1^d (r) \sim r^{k-1}$, and
$V_1^d (r) \sim r^\sigma$ as $r \to 0$, we may always match these
solutions of the far region in the overlap region by adjusting $u^r_0$ and $c_+$
in (\ref{61b}),  and $v_+$ in (\ref{61c}) without creating any relations
between the sources and the expectation values. Finally, since $V_1 ^d (\infty)=1$,
the coefficient $\zeta _d$ is given by the source $g^{(0)} _{ii}$ for $i=1,2$.
As a result, the expectation values of the operators $\cT_{--}$, $\cT_{+-}$, $\cT_{++}$,
$\cJ_+$, $\cJ_-$, and $\cT_K$ are all local functions of $g^{(0)}_{ii}$. The
corresponding correlators vanish, and by locality of the total trace
$\cT$, so do the correlators with the combination $\cT^{+-} + \beta \cT^{++}$,
and we have,
\bea
\left \< \cJ_\pm (p) \, \cT_K (-p) \right \>
= ~ \left \< \cJ_\pm (p) \, \left ( \cT_{+-} - \beta \cT_{--} \right )(-p)  \right \> & = & 0
\no \\
\left \< \cT_{ \pm \pm }(p) \, \cT_K (-p) \right \> =
\left \< \cT_{ \pm \pm }(p) \,  \left ( \cT_{+-} - \beta \cT_{--} \right )(-p) \right \> & = & 0
\no \\
\left \< \cT_{ +- }(p) \, \cT_K (-p) \right \>
= \left \< \cT_{ +- }(p) \,  \left ( \cT_{+-} - \beta \cT_{--} \right )(-p) \right \>  & = & 0
\no \\
\left \< \cT_K (p) \, \cT_K (-p) \right \>
= ~ \left \< \cT_K (p) \,  \left ( \cT_{+-} - \beta \cT_{--} \right )(-p) \right \>  & = & 0
\eea
These results form the basis for the derivation of the relations (\ref{91n}).

\subsection{Expansion of $N_1$ in powers of $q_+$}

The solution for the field $N_1(r)$ was constructed in (\ref{63c}),
\bea
\label{53a}
N_1(r)= {L_0(r)  \over 2b}  s_{--} - 2 b  L_0^c (r) t_{--}
\eea
The functions $L_0(r)$ and $L_0^c(r)$ are data of the purely magnetic background
solution, and thus independent of $q_+$. Therefore, all $q_+$ dependence is
contained in the coefficients $s_{--}$ and $t_{--}$, which themselves admit a Taylor
expansion in $q_+$. In fact, a Taylor expansion holds for all
components of the tensors $s_{\mu \nu}$ and $t_{\mu \nu}$, and we
have,\footnote{The order of expansion in powers of $q_+$
will be denoted by a square bracket superscript, while the Fefferman-Graham order of expansion in powers of
$\rho^{-1}$ will be  denoted by superscript parentheses.}
\bea
\label{53c}
s_{\mu \nu} & = & s_{\mu \nu}^{[0]} +  s_{\mu \nu}^{[1]} +  s_{\mu \nu}^{[2]} + \cdots
\no \\
t_{\mu \nu} & = & t_{\mu \nu}^{[0]} +  t_{\mu \nu}^{[1]} +  t_{\mu \nu}^{[2]} + \cdots
\eea
It is important to notice that, even though $N_1$ is a metric fluctuation, its
expansion in powers of $q_+$ may contain odd powers through the gauge field sources.

\sm

The relations between $s_{\pm \pm}$ and $t_{\pm \pm}$, given
on the first two lines of (\ref{61b}) may similarly be expanded in powers of $q_+$.
Since the coefficients of these relations are independent of $q_+$,
the relations are found to hold order by order,
\bea
\label{53d}
t_{++}^{[n]}  & = & { p_+ p_- \over 24 b} s_{++}^{[n]} +{ p_+^3 \over 24 b p_-} s_{--} ^{[n]}
\no \\
t_{--} ^{[n]}  & = & { p_+ p_- \over 24 b} s_{--} ^{[n]}+{ p_-^3 \over 24 b p_+} s_{++} ^{[n]}
\eea
Next, we use the key observation of the end of section \ref{sixone}: the sources
$\go_{\mu \nu}$ and $\Ao_\mu$ are independent of $q_+$. In particular, the
source of the field $N_1$ must be independent of $q_+$. But this source may
be identified from the $r \to \infty$ asymptotics of the solution by  (\ref{52a})
and, using $L_0(r) \sim 2br$ and $L_0^c (r) \sim -1/(2b)$ is found to be given by,
$4 \go_{--} = s_{--}$. As a result we have,
\bea
\label{53e}
s_{--} ^{[0]} & = & {1 \over 4b} \go_{--}
\no \\
s_{--} ^{[n]} & = & 0 \hskip 1in n\geq 1
\eea
and thus (\ref{53d}) simplifies as follows,
\bea
\label{53f}
t_{++}^{[n]}  & = & { p_+ p_- \over 24 b} s_{++}^{[n]} \hskip 1in n\geq 1
\no \\
t_{--} ^{[n]} & = & { p_-^3 \over 24 b p_+} s_{++} ^{[n]}  \hskip 1in n\geq 1
\eea
which implies $p_+^2 t_{--} ^{[n]} = p_-^2 t_{++}^{[n]} $.
The $r \to \infty$ asymptotics allows us to extract the
expectation value of the normalized stress tensor component $T_{--}$
and we find,
\bea
\label{53f}
16 \pi G_3 T_{--}= c_V g^{(4)} _{--} = 2 b  \, t_{--}
\eea
Note that  all (even) orders in $q_+$ are allowed to contribute to $t_{--}$.

\subsection{Expansion of $A_-$ in powers of $q_+$}

The solution for $A_-$ obtained in (\ref{63d}), with the overlap region
matching conditions of (\ref{63f}),  is given by,
\bea
\label{54a}
A_-(r) = \s_-  + { A_0(r) \over 4b}s_{--} - b  {A_0L_0^c \over L_0}(r) t_{--}
-   {M_0 \over 4k L_0 A_0}t_{--}
\eea
and also admits a Taylor expansion in powers of $q_+$.
Let us now investigate how this expansion affects
the source and vev for this field. To derive the $r \to \infty$ asymptotics
of $A_-$, we make use of the fact that $L_0^c/L_0 =\cO(r^{-2})$,
as well as the following auxiliary asymptotics,
\bea
\label{54c}
A_0(r) & \sim & {c_V c_E \over kb} \left ( 1 -{ kb \over c_V r} \right ) + \cO(r^{-2})
\no \\
{M_0(r) \over L_0(r)} & \sim & - {\a_c \over 2b} +\cO(r^{-2})
\eea
As a result, we find the following equations for the source term $\Ao_-$,
and the vev term $\At_-$,
\bea
\label{54d}
\Ao_- & = & \s _-  +{c_V c_E \over 12k} s_{--}
+ {\a_c \over 8 c_V c_E} t_{--}
\no \\
\At_- & = &   - { c_E \over b} s_{--}
+ {\a_c kb \over 2 c_V^2 c_E} t_{--}
\eea
Using the fact that the source $\Ao_-$ is independent of $q_+$,
and the earlier result (\ref{53e}), we obtain the following
relations between the expansion coefficients in powers of $q_+$,
\bea
\label{54e}
\Ao_- & = & \s _-  ^{[0]}
\no \\
0 & = & \s _-^{[1]}  +{c_V c_E \over 48k} \go_{--}
+ {\a_c \over 8 c_V c_E} t_{--} ^{[0]}
\no \\
0 & = & \s _- ^{[n]}  + {\a_c \over 8 c_V c_E} t_{--}^{[n-1 ]}
\hskip 1in n \geq 2
\eea
where we have used the fact that $c_E$ and $\a_c / c_E$ are linear in $q_+$.
The expectation value of the normalized current component $\cJ_-$ is given by,
\bea
\label{54f}
8 \pi G_3 J_- = c_V \At_- =
 {\a_c kb \over 2 c_V c_E} \, t_{--}
\eea
where we have omitted the local term proportional to $s_{--}$ in deriving
this result from (\ref{54d}).

\subsection{Proportionality of $\cJ_-$ and $\cT_{--}$}
\label{JT}

Comparing the formulas for the expectation values for $\cJ_-$ in
(\ref{54f}) and for $\cT_{--}$ in (\ref{53f}), we readily obtain a simple
identity between these expectation values,
\bea
\label{56a}
J_-  =  Q_+  T_{--}
\hskip 1in  Q_+ = { k \a_c \over 2 c_V c_E}
\eea
In deriving this result, we have omitted the fractional power dependence
on $p^2$ by setting $c_-=v_-=0$. Since the result is obtained by overlapping
expansion methods, we are of course also omitting the integer power
corrections in $p^2$ and $q_+p_-$. In sum, relation (\ref{56a}) holds
within the approximations of section \ref{approx}. Since (\ref{56a})
holds for any sources, it is equivalent to the operator identity,
$\cJ_-  \approx  Q_+  \cT_{--} $ announced in (\ref{93a}). This equation is invariant
under extended Lorentz symmetry, since $Q_+ $ is linear in $q_+$.

\newpage

\section{Calculation of the correlators of $\cJ_+$ and $\cT_{++}$}
\setcounter{equation}{0}
\label{seven}

Using the general set-up of the preceding section, all correlators involving
$\cJ_+$ and $\cT_{++}$ will be evaluated in this section. The calculations will
be organized along the lines laid out in section \ref{four}, and within the
confines of the approximations discussed in section \ref{approx}. Correlators
involving $\cJ_+$ will be reached by turning on the single source $A^{(0)}_-$,
while leaving the remaining  sources $A_+^{(0)}$ and $g^{(0)}_{\pm \pm}$
turned off. Similarly, correlators involving $\cT_{++}$ will be reached by
turning on only the source $g^{(0)}_{--}$. The chiral correlators $\< \cJ_+ \cJ_+\>$,
$\< \cJ_+ \cT_{++}\>$, and $\< \cT_{++} \cT_{++}\>$ will be evaluated in this
manner to all orders in $q_+$.  We will also take the opportunity to check
that correlators of $\cJ_+$ and $\cT_{++}$ involving mixed chiralities, or the
trace of the stress tensor, vanish as announced in (\ref{91n}) and (\ref{93b}).

\subsection{Sourcing $\cJ^-$}

We begin by sourcing only the current $\cJ^-$ by turning on $A_- ^{(0)}$.
This case is the simplest, and we will discuss the corresponding calculation in detail.
The general solution for the reduced Maxwell fields $a_\pm(r)$ in the overlap
region was derived in (\ref{61b}). Using (\ref{63f}), it may
alternatively be expressed as follows,
\bea
\label{57a}
a_+ (r) & = & \sigma _+ +  {k \tau _+ \over q_+} A_0(r)
\no \\
a_- (r) & = & \sigma _-
\eea
Here, we have used the $r\to 0$ asymptotics of the function $A_0 (r)$ in (\ref{23c})
to recast (\ref{61a}) in terms of $A_0$ in the overlap region.
The corresponding  Maxwell field strengths are given by,
\bea
\label{57b}
E_1(r) & = & {k \tau _+ \over q_+} E_0(r)
\no \\
P_1(r) & = & 0
\eea
We will now show that the form of the perturbations exhibited in (\ref{57a}) and
(\ref{57b}), although initially derived for, and valid in, the overlap region only,
actually provides an exact solution valid throughout the far region,
under certain conditions. To establish this, we point out that the vanishing
of $P_1$, together with the proportionality of $E_1(r)$ to $E_0(r)$, indicates that
these perturbations may be viewed as the result of transforming $q_+$ by,
\bea
\label{57c}
q_+ & \to & q_+ + k \tau_+
\eea
Since $\tau_+$ should be viewed as a linear perturbation, this shift
is infinitesimal. It may be interpreted as a rescaling of $q_+$ and
may be achieved by an $SL(2,\bR)$ transformation of (\ref{27b}) with
$\lambda _0 = 1+k \tau_+/q_+$ and $\lambda =0$.
Here, $\tau_+$ actually depends on momenta and on source functions, but
from the point of view of the differential equations in $r$ for the far
region, this extra dependence is inconsequential.

\subsubsection{Solution in the far region}

The coefficients $\s_\pm$ and $\tau_+$ may be eliminated in terms of the
sources\footnote{It will be instructive to temporarily keep both sources
$A_\pm^{(0)}$ even though we will ultimately set $A^{(0)}_+=0$.}
$A_\pm ^{(0)}$ to the Maxwell field $A_\pm$. Using the large
$r$ asymptotics result for $A_0 (r)$ given in (\ref{23a}), and the relation
$p_+ \s_- = p_- \s_+$ derived from (\ref{61b}), we find,
\bea
\label{57d}
a_+ (r) & = & {p_+ \over p_-} A^{(0)} _- +
\left ( A_+ ^{(0)} - { p_+ \over p_-} A_- ^{(0)} \right ) { kb \over c_V c_E} A_0(r)
\no \\
a_- (r) & = & A_- ^{(0)}
\eea
Since the solution (\ref{57d}) is effectively generated by the transformation of (\ref{57c}),
we readily determine the solutions in the far region for the remaining functions,
and we find,
\bea
\label{57e}
0 & = & L_1 = N_1 = P_1 =V_1=f_1
\no \\
M_1(r) & = & m_0 L_0(r) + m_1 L_0^c (r) + {2kb \over c_V c_E}
\left ( A_+ ^{(0)} - { p_+ \over p_-} A_- ^{(0)} \right ) M_0(r)
\eea
The first two terms in $M_1$ provide homogeneous solutions to the
equation E2 for $M_1(r)$, while the third term provides an inhomogeneous
solution sourced by the perturbation $E_1(r)$.
%
%
In partial summary, the fields of (\ref{57d}) and (\ref{57e}) solve the
far region equations, and their Maxwell fields properly match onto those
of the overlap solution. To make them into full fledged solutions, it remains
to ensure that also the metric functions match with those of the solution in
the overlap region, and satisfy the absence of metric sources as $r \to \infty$.

\subsubsection{Matching the metric solutions in the overlap region}
\label{sevenonetwo}

Matching with the overlap region solution determines the coefficients,
\bea
\label{57f}
m_0 = {s_{++} \over 2b} \hskip 1in m_1 = - 2 b t_{++}
\eea
and insisting on vanishing metric sources as $r \to \infty$  gives the following relations,
\bea
\label{57g}
s_{--}^{[n]} & = & 0  \hskip 2.5in  n \geq 0
\no \\
s_{++} ^{[n]} & = &  s_{++} ^{[0]} =0  \hskip 2in n \geq 2
\no \\
s_{++} ^{[1]} & = & { 2kb \a_c \over c_V c_E}
\left ( A_+ ^{(0)} - { p_+ \over p_-} A_- ^{(0)} \right )
\eea
With the help of  these equations, we evaluate $t_{\pm \pm} ^{[n]}$ using the
overlap region solution of (\ref{53d}) and (\ref{53e}) for $g^{(0)}_{--}=0$,
and we find,
\bea
\label{57h}
t_{++}^{[n]} & = & t_{--}^{[n]} =0 \hskip 2.5in  n \geq 2
\no \\
t_{++}^{[0]} & = & t_{--}^{[0]} =0
\no \\
t_{++}^{[1]} & = & { 2 kb \a_c \over c_V c_E} \, {p_+ p_- \over 24b}
\left ( A_+ ^{(0)} - { p_+ \over p_-} A_- ^{(0)} \right )
\no \\
t_{--}^{[1]} & = & { 2 kb \a_c \over c_V c_E} \, {p_-^3 \over 24 b p_+}
\left ( A_+ ^{(0)} - { p_+ \over p_-} A_- ^{(0)} \right )
\eea
Clearly, $t_{++}^{[1]}$ will provide only local terms in the expectation
value $T_{++}$ and may be omitted within our approximation.
The term sourced by $A_- ^{(0)}$ in $t_{--}^{[1]}$  similarly
produces only local terms and may be omitted. Thus, if we set $A_+^{(0)}=0$,
then all contributions to $t_{\pm \pm}$ are local and may be
omitted in the matching process in the overlap region.
In conclusion, when $A_+^{(0)}=0$, the expressions of (\ref{57d})
and (\ref{57e}) provide an exact solution to all orders in $q_+$,
and match with the solution in the overlap region, up to local terms.

\sm

To close, it is interesting to see what happens when $A_+ ^{(0)} \not= 0$.
There is now a non-local contribution to $N_1$, forced upon us by the
matching of the solution in the far region with the solution in the overlap
region. But this contribution, in turn, will source $P_1$ as well as $V_1, f_1, L_1, E_1$
and $M_1$. Therefore,  when $A_+ ^{(0)} \not= 0$, the
expressions of (\ref{57d}) and (\ref{57e}) are {\sl incompatible} with the
overlap solution, even within  the approximations of section \ref{approx}.

\subsubsection{Calculation of correlators with $\cJ^-$}

Putting together the result of section \ref{sevenonetwo},
under the assumption that $A_+^{(0)}=0$, we obtain the following solution,
\bea
\label{57i}
a_+ (r) & = & {p_+ \over p_-} A^{(0)} _- \left ( 1
 - { kb \over c_V c_E}   A_0(r) \right )
\no \\
a_- (r) & = & A_- ^{(0)}
\no \\
M_1(r) & = &
- {2kb \over c_V c_E} { p_+ \over p_-} A_- ^{(0)}
\left ( M_0(r) + {\a_c \over 2b} L_0(r) \right )
\eea
with the remaining fields $L_1=N_1=V_1=0$, up to local terms. This
solution is valid to all orders in $q_+$, within the approximation of
section \ref{approx}. Using the $r \to \infty$ asymptotics of  $A_0$ given in (\ref{23a}),
and the following $r \to \infty$ asymptotics for $M_0$,
\bea
\label{57j}
M_0(r) +{ \a_c \over 2b} L_0(r) = ~ \sim ~ {c_V c_E^2 \over kb r}
\eea
we derive the asymptotics of the fields as $r \to \infty$, and we find,
\bea
\label{57k}
a_+ (r) & \sim  & {p_+ \over p_-} A^{(0)} _- \, {kb \over c_V r}
\no \\
a_- (r) & = & A_- ^{(0)}
\no \\
M_1(r) & = &
- { p_+ \over p_-} A_- ^{(0)} \, { 2c_E \over  r}
\eea
The expectation values of the current $\cJ^-$ with all components of
$\cJ$ and $\cT$ may be read off from (\ref{57k}) and $L_1=N_1=V_1=0$.
First of all, a number of expectation values are found to vanish,
$J_- = T_{--} = T_{+-} =T_K= 0$, within the approximations of section \ref{approx}.
This result implies that the corresponding correlators also vanish,
\bea
\label{71a}
\left \< \cJ_- (p) \, \cJ^{-} (- p) \right \> & = & 0
\no \\
\left \< \cT_{--} (p) \, \cJ^{-} (- p) \right \> & = & 0
\no \\
\left \< \cT_{+-} (p) \, \cJ^{-} (- p) \right \>  & = & 0
\no \\
\left \< \cT_K (p) \, \cJ^{-} (- p) \right \>  & = & 0
\eea
Converting the upper index to lower indices,
\bea
\label{71b}
\cJ^- =  2 \cJ_+ - 4 \beta \cJ_-
\eea
we see that the results of (\ref{71a}) are all consistent
with the correlators (\ref{93b}) predicted on general grounds.

\sm

The remaining correlators with $\cJ^-$ are  governed by
the following expectation values,
\bea
\label{57l}
J_+ & = & {c_V \over 8 \pi G_3} A^{(2)} _+
= {kb \over 2 \pi G_3} {p_+ \over p_-} A_-^{(0)}
\no \\
T_{++} & = & {c_V \over 16 \pi G_3} g^{(4)} _{++}
= -  { c_V c_E \over 2 \pi G_3} {p_+ \over p_-} A_-^{(0)}
\eea
As a result, we find the corresponding correlators,
\bea
\label{57m}
\< \cJ_+ (p) \, \cJ^-(-p) \>
& = & {kb \over 2 \pi G_3} \, {p_+ \over p_-}
\no \\
\< \cT_{++} (p) \, \cJ^- (-p) \>
& = & - {  c_V c_E \over 2 \pi G_3} \, {p_+ \over p_-}
\eea
Using again (\ref{71b}) and the  vanishing of the correlators of mixed chirality,
we derive,
\bea
\label{71d}
\< \cJ_+ (p) \, \cJ_+(-p) \> & = & {kb \over 4 \pi G_3} \, {p_+ \over p_-}
\no \\
\< \cT_{++} (p) \, \cJ_+ (-p) \> & = & - {   c_V c_E \over 4 \pi G_3} \, {p_+ \over p_-}
\eea
These results hold to all orders in $q_+$. The coefficient on the first line
coincides with the one computed for the purely magnetic solution (since it is
independent of $q_+$), while the second coefficient is the one announced
in (\ref{92e}) and (\ref{92f}).

\subsection{Sourcing only $\cT^{--}$}

We source $\cT^{--}$ by turning on $g^{(0)} _{--}$ while keeping
$g_{+-} ^{(0)} =g_{++}^{(0)}=g_{ii}^{(0)}=A^{(0)} _\pm=0$. In view of (\ref{53a}), the field $N_1$ is
turned on since $s_{--}=s_{--}^{[0]} = 4b g^{(0)} _{--}$.
One might now solve successively the fluctuation equations of (\ref{62d})
and (\ref{62e}) following the scheme of (\ref{63aa}). Actually there is a
simpler and more illuminating way of proceeding using the exact $SL(2,\bR)$
invariance of the reduced field equations in the far region, given in (\ref{21dd}).
In the far region, the source $g_{--}^{(0)}$ may be turned on by
performing the following $SL(2,\bR)$ transformation $\Lambda$ on the fields,
\bea
\label{72a}
\Lambda = \left ( \matrix{ 1 &  \ep g_{--}^{(0)} e^{ipx} \cr & \cr 0 & 1 \cr} \right )
\eea
Using the transformation rules of (\ref{21dd}) on the full fields
$L,M,N,E,P,V$ of (\ref{62a}) leads to the following exact solution
to the fluctuation equations in the far region,
\bea
\label{72b}
M_1 (r) = 0 & \hskip 1in & L_1(r) = g_{--}^{(0)} M_0(r)
\no \\
E_1(r) = 0 && N_1(r) = 2 g_{--}^{(0)} L_0(r)
\no \\
V_1(r) = 0 && P_1(r) = - g_{--}^{(0)} E_0(r)
\eea
This far region solution is incompatible with the matching conditions in the
overlap region, however,  because the non-zero source
$s_{--}^{[0]} = 4b g^{(0)} _{--} \not= 0$ implies that $t_{++}^{[0]} \not= 0$
is non-local, in view of (\ref{53d}). Therefore it cannot be consistently omitted in our calculations.
This shortcoming is easily remedied by adding to $M_1$ the homogeneous
solutions $L_0$ and $L_0^c$, with coefficients to be determined
by the matching in the overlap region. We will do so below.

\sm

Since the above solution has a non-trivial $P_1$-field,
we are inadvertently turning on an unwanted source in the field $A_-$.
From the preceding section, we know
how to turn this source off by a shift in $q_+$ as in (\ref{57c}).
Since the fluctuation equations are linear, the two solutions may simply be
added in the far region. The resulting total solution takes the form,
\bea
\label{72cc}
L_1(r) & = & g_{--}^{(0)} M_0(r)
\no \\
M_1(r) & = & {s_{++} \over 2b} L_0(r) -2b t_{++} L_0^c(r) + 2 \tau M_0(r)
\no \\
N_1(r) & = & 2 g_{--}^{(0)} L_0(r)
\no \\
a_+ (r) & = & \sigma _+ + \tau A_0(r) \hskip 1.65in E_1= \tau E_0
\no \\
a_- (r) & = & \sigma _- + g^{(0)} _{--} A_0(r)  \hskip 1.5in P_1= - g^{(0)}_{--}  E_0
\eea
while we continue to have $V_1=0$.
This solution holds to all orders in $q_+$ as may readily be checked by
explicit calculation.

\subsubsection{Absence of sources and matching the solution in the overlap region}

The absence of the sources $g_{++}^{(0)}$ and $A_\pm ^{(0)}$,
and the presence of the source $g_{--}^{(0)}$ imposes the following relations
on the metric coefficients,
\bea
\label{72c}
s_{++}  & = & 2  \a_c \tau
\no \\
s_{--} ^{[0]} & = & 4b g^{[0]} _{--}
\no \\
s_{--} ^{[n]} & = & 0 \hskip .5in  n \geq 1
\eea
and the relations
\bea
\label{72d}
0 & = & \sigma _+ + \tau { c_V c_E \over kb}
\no \\
0 & = & \sigma _- + g^{(0)} _{--}  { c_V c_E \over kb}
\eea
on the Maxwell coefficients. We solve (\ref{72d}) for $\tau$ by exploiting the
overlap region relation $p_+ \sigma_- - p_- \sigma _+=0$,
then use this result for $\tau$ to solve (\ref{72c}) for $s_{++}$, and we find,
\bea
\label{72e}
\tau =  {p_+ \over p_-} g^{(0)} _{--} \hskip 1in
s_{++} = 2 \a_c \, {p_+ \over p_-} \, g^{(0)} _{--}
\eea
The second relation provides $s_{++}$ with a simple expansion in powers of $q_+$
given by,
\bea
\label{72f}
s_{++}^{[0]} & = & s_{++}^{[1]} = 0
\no \\
s_{++}^{[2]} & = & 2  \a_c \, {p_+ \over p_-} \, g^{(0)} _{--}
\no \\
s_{++}^{[n]} & = & 0  \hskip 1in n \geq 3
\eea
Using the remaining equations of (\ref{72c}), and the matching
conditions for the metric (\ref{53d}),
\bea
\label{72g}
t^{[0]} _{++} = {p_+^3 \over 24bp_- } s_{--}^{[0]}
& \hskip 0.6in  &
t^{[2]} _{++} = {p_+ p_- \over 24b} s_{++}^{[2]}
\hskip 0.6in
t^{[1]} _{++} = t^{[1]} _{--} =0
\\
t^{[0]} _{--} = {p_+p_- \over 24b} s_{--}^{[0]} ~
& \hskip 0.6in &
t^{[2]} _{--} = {p_-^3  \over 24bp_+ } s_{++}^{[2]}
\hskip 0.6in
t^{[n]} _{++} = t^{[n]} _{--} =0 \hskip 0.5in n \geq 3
\no \eea
Substitution of $s^{[2]}_{++}$, given by (\ref{72f}) into the expressions
for $t_{\pm \pm} ^{[2]}$ immediately reveals that these quantities are
local, and so of course is $t_{--} ^{[0]}$. Retaining only source terms and
non-local contributions in the  fields gives $V_1=0$ and,
\bea
\label{72h}
L_1(r) & = & g_{--}^{(0)} M_0(r)
\no \\
M_1(r) & = & 2\,  {p_+ \over p_-} g^{(0)} _{--} \Big (  M_0(r)
+ {\a_c \over 2b} \,  L_0(r) \Big )
- {p_+^3 \over b p_-} g^{(0)}_{--}L_0^c(r)
\no \\
N_1 (r) & = & 2 g_{--} ^{(0)} L_0(r)
\no \\
a_+(r) & = & {p_+ \over p_-} g^{(0)} _{--} \left ( A_0(r) - {c_V c_E \over kb} \right )
\no \\
a_- (r) & = & g_{--} ^{(0)} \left ( A_0 (r) - { c_V c_E \over kb} \right )
\eea
This solution holds to all orders in $q_+$, within the approximations of
section \ref{approx}. Extracting the leading asymptotics, it is now verified
that only the fluctuations $L_1$ and $N_1$ have a source,
while $M_1$, and $a_\pm$ do not.

\sm

The presence of a linear term in $L_1(r)$ as $r \to \infty$ implies that we
have turned on a source to the operator $\cT^{+-}$ as well as to $\cT^{--}$.
More precisely, we have introduced the coupling of $g^{(0)} _{--}$ to
the combination
\bea
\cT^{--} - 2 \beta \cT^{+-} = 4 \cT_{++} + 8 \beta ^2 \cT_{--}
\eea
From the sub-leading asymptotics in (\ref{72h}), we derive the non-local parts of the
expectation values. We clearly find $J_-=T_{+-}=T_{--}=T_K=0$,
and these relations imply the vanishing of the following correlators,
\bea
\label{72j}
\left \< \cJ_- (p) \, ( \cT_{++} + 2 \beta ^2 \cT_{--})  (-p) \right \> & = & 0
\no \\
\left \< \cT_{+-} (p) \, ( \cT_{++} + 2 \beta ^2 \cT_{--}) (-p) \right \> & = & 0
\no \\
\left \< \cT_{--} (p) \, ( \cT_{++} + 2 \beta ^2 \cT_{--}) (-p) \right \> & = & 0
\no \\
\left \< \cT_K (p) \, ( \cT_{++} + 2 \beta ^2 \cT_{--}) (-p) \right \> & = & 0
\eea
These relations are consistent with the relations (\ref{91n}) and (\ref{93b})
derived on general grounds.

\sm

The remaining expectation values are non-vanishing, and given by,
\bea
\label{72i}
g^{(4)}_{++} & = &
{p_+^3 \over b \, c_V \, p_- } g^{(0)} _{--} + { 8 c_V c_E^2 \, p_+ \over kb \, p_-} g^{(0)} _{--}
\no \\
A^{(2)} _+ & = & - { 4 c_E \, p_+ \over p_-} g^{(0)} _{--}
\eea
whence we derive the expectation values of the normalized current and stress tensor,
\bea
\label{72j}
J_+ & = & - { c_V c_E p_+ \over 2 \pi G_3 \, p_-} \, g^{(0)} _{--}
\no \\
T_{++} & = & {p_+^3 \over 16 \pi b  G_3 \, p_-} g^{(0)} _{--}
+ {  c_V^2 c_E^2 \, p_+  \over 2 \pi G_3 kb \, p_- }   g^{(0)} _{--}
\eea
These relations hold to all orders in $q_+$. We readily deduce the
correlators in momentum space, and find,
\bea
\label{72k}
\left \< \cJ _+ (p) \, ( \cT_{++} + 2 \beta^2 \cT_{--} ) (-p) \right \>
& = & - {  c_V c_E p_+ \over 4 \pi G_3 \, p_-}
\no \\
\left \< \cT_{++} (p) \, ( \cT_{++} + 2 \beta^2 \cT_{--} ) (-p) \right \> & = &
{p_+^3 \over 32 \pi b \, G_3 \, p_-}
+ {   c_V^2 c_E^2 \, p_+  \over 4 \pi G_3 kb \, p_- }
\eea
Using the vanishing of the corresponding mixed chirality correlators
from (\ref{91n}) and (\ref{93b}), we obtain our final result,
\bea
\label{72l}
\left \< \cJ_+ (p) \cT_{++} (-p) \right \> & = & - {   c_V c_E p_+ \over 4 \pi G_3 \, p_-}
\no \\
\left \< \cT_{++} (p) \cT_{++} (-p) \right \> & = &
{p_+^3 \over 32 \pi b \, G_3 \, p_-}
+ {   c_V^2 c_E^2 \, p_+  \over 4 \pi G_3 kb \, p_- }
\eea 
The result on the first line agrees with the first line of (\ref{57m}): this
gives a check that a correlator which may be evaluated in two different
ways does indeed come out uniquely. The first term on the second line
reproduces the purely magnetic result on the second line of (\ref{91a}),
while the second term gives the result announced in (\ref{92a}) and (\ref{92f}).
 
\newpage
\section{Discussion}
\setcounter{equation}{0}
\label{eight}

The purpose of this work was to compute long-distance correlation functions in the
charged magnetic background solution as a means of probing the low-energy
dynamics of the dual system of fermions at finite density and magnetic field.
The results obtained  shed light on some properties of this theory, but various
questions remain.

\sm

In particular, the existence of a quantum critical point in this system is now well
established, both numerically and analytically,
but we lack a good understanding of what is driving this transition.
Since we are working with a fully top-down construction in which the
dual gauge theories are completely specified (e.g., ${\cal N}$=4 SYM),
we can in principle hope to obtain complementary descriptions
of the transition mechanism within both gauge theory and gravity.
It would be very instructive to have such an understanding of a
finite density quantum phase transition from these two points of view.
Note that in contrast to models of quantum criticality involving probe
fermions  \cite{Liu:2009dm,Cubrovic:2009ye,Faulkner:2009wj,Faulkner:2010tq},
here the gauge theory fermions are expected to be important players in
the dynamics of the phase transition, rather than spectators.

\sm

The results obtained here for the correlators of the stress tensor and U(1)
current provide new information about the nature of the critical point.
At least as probed by these operators, the low energy
theory appears to retain the character of a 1+1 dimensional CFT,
albeit with twisted Virasoro generators and renormalized propagation speed.
This behavior seems consistent with a general 1+1 dimensional Luttinger
liquid description.  On the other hand,  what is puzzling is that this gives no
hint as to why the low temperature thermodynamics exhibits the nontrivial
specific heat behavior displayed in equation (\ref{1b}) of the Introduction;
such behavior is  not characteristic of Luttinger liquid behavior, but rather
indicates the breakdown of such a description.

\sm

To appreciate this point, it is useful to compare the situation found here to
one in which the low energy effective theory is Lorentz invariant.  In a Lorentz
and scale invariant theory with a traceless stress tensor, it is simple to show
that there is a direct relation between the stress tensor correlator and the
specific heat exponent; the power law obeyed by the former determines the latter.
For instance, a $1/x^4$ falloff of the stress tensor correlator in 1+1-dimensions
fixes the specific heat to be linear in temperature.   In our case, we also have
a 1+1-dimensional effective theory with a stress tensor correlator exhibiting
$1/x^4$ falloff, but the specific heat is not, in general, linear in the temperature.
This can be traced back to the lack of Lorentz invariance; specifically, due to
the fact that the stress tensor cannot be made to be both symmetric and
traceless (see  \cite{Guica:2010sw} for a recent  discussion of this point).
The conclusion is that in a non-Lorentz invariant theory one cannot
necessarily expect to see nontrivial specific heat behavior mirroring itself in
the behavior of the stress tensor correlator, which is consistent with our computations.

\sm

The question thus remains as to what other observable might exhibit
qualitative features sensitive to the degrees of freedom that are going
critical at the quantum phase transition.  One natural possibility is to
consider correlation functions of charged fields in the bulk.  We expect
that these can be computed using the same methods employed in this paper.
Ideally, such correlators will exhibit behavior at the critical point that can be
used to elucidate the mechanism driving the transition in the dual gauge theory.
We leave these investigations for future work.

\newpage

\appendix

\section{Scaling laws for low temperature thermodynamics }
\setcounter{equation}{0}
\label{appAA}

In this appendix we review the $k$-dependent behavior of the low temperature
thermodynamics around the quantum critical point.  Recalling that we can,
without loss of generality, assume $k\geq 0$, there turn out to be three distinct
regions :   $0 \leq k < 1/2$;  $1/2 \leq k \leq 3/4$;  and $k \geq 3/4$.

\sm

We recall that for our system the quantum critical point arises at
a particular value of the dimensionless magnetic field, $\hat{B}= \hat{B}_c$,
obeying the following properties.  For $\hat{B} < \hat{B}_c$ the entropy density
remains finite at zero temperature; for $\hat{B} > \hat{B}_c$ the entropy density
goes to zero linearly with temperature; and for $\hat{B} = \hat{B}_c$ the entropy density
vanishes as a nontrivial power law, $\hat{s} \sim T^\alpha$.  The
values of $\hat{B}_c$ and $\alpha$ are $k$-dependent quantities in general.

\sm

For $k< 1/2$ there is no quantum critical point, as the zero temperature entropy
density is found to be nonzero for any value of $\hat{B}$.  This was established
numerically in \cite{D'Hoker:2010rz}.  The special role
of $k=1/2$ can be seen by examining the near horizon solution (\ref{23b}).
This solution clearly breaks down at $k=1/2$, and for $k<1/2$ has  the wrong
large $r$ behavior of $g_{++}$ to match onto an asymptotically AdS$_5$ geometry.
This near horizon geometry thus plays no role for $k<1/2$, and the low temperature
behavior is instead governed by a finite entropy density solution whose form has not been
determined analytically as yet.

\sm

Next consider the range $1/2 \leq k \leq 3/4$.  The low temperature
thermodynamics in this region is found to be controlled by the near horizon solution
(\ref{23b}).  This is a scale invariant geometry, and as discussed in section \ref{dynam},
the thermodynamic behavior associated with this scaling is:
\bea
\label{37ee}
\hat s \sim \hat T ^\alpha \hskip 1in \alpha = {1-k \over k}
\eea
In Table \ref{alph} we compare this prediction against numerical data, and the
agreement is seen to be excellent.
\begin{table}[htdp]
\begin{center}
\begin{tabular}{|c||c|c|c|} \hline
$k$ & $\alpha$   & $(1-k) / k$ & $\Delta \alpha$ \\ \hline \hline
0.55 & 0.8182   & 0.8182 & 0.0003 \\ \hline
0.60 & 0.6667   & 0.6667 & 0.0003 \\ \hline
0.65 & 0.5389   & 0.5385 & 0.0003 \\ \hline
0.70 & 0.4319   & 0.4286 & 0.0026 \\ \hline
0.749 & 0.3606 & 0.3351 & 0.0032 \\ \hline \hline
\end{tabular}
\end{center}
\caption{Numerical results for the exponent $\alpha$,  its statistical standard
deviation $\Delta \alpha$, and comparison with the analytically predicted values
$(1-k)/k$, in the scaling relation $\hat s \sim \hat T ^\alpha$ for the range $1/2 < k < 3/4$.}
\label{alph}
\end{table}
As we have noted, this scaling behavior only holds in the range
$1/2 \leq k \leq 3/4$, and we now discuss why this is the case.  In order for a
given near horizon, zero temperature solution, such as (\ref{23b}),
to control the low temperature thermodynamics, it must be the case
that there exists a finite temperature solution that shares the same
asymptotics.  Only in this case will the scaling behavior of the zero temperature
solution manifest itself at finite temperature.  What happens for $k > 3/4$ is that
at finite temperature the  function $V$ appearing in the general metric Ansatz
(\ref{21a}) turns out to grow as one moves out from the horizon.  This is
incompatible with the asymptotics of (\ref{23b}) in which $V$ is constant.
The fact that this behavior sets in at $k=3/4$ can be seen numerically,
but a simple analytical explanation is presently lacking.

In the range $k>3/4$ the low temperature thermodynamics is not associated
with the scaling behavior of a near horizon solution, but
rather emerges from the more involved asymptotic matching analysis performed
in \cite{D'Hoker:2010ij}.   This analysis, which confirms the earlier numerical
results presented in \cite{D'Hoker:2010rz},
leads to the scaling behavior $\hat{s} \sim \hat{T}^{1/3}$.  Note
that this agrees with the value of $\alpha$ quoted in (\ref{37ee}) for $k=3/4$.

\section{Perturbation theory and WKB for scalar correlators}
\setcounter{equation}{0}
\label{appA}

In this appendix, we derive a systematic perturbation theory in powers of
small $\xi$ for the scalar correlators of section \ref{three}, and develop a WKB
approximation for large $\xi$.

\subsection{Perturbative expansion in power of $\xi$}

For $1/2<k$ the $z^{4-4k}$ term in equation (\ref{35e}) is suppressed compared to the $z^2$ term
for large $z$. For small $z$, this situation is reversed, but the $\nu^2$ term then dominates
over both  $z^2$ and $z^{4-4k}$, as long as $k <1$. Thus, in the interval $1/2 < k < 1$,
one expects a convergent perturbative expansion in the $z^{4-4k}$ term to hold.
The starting point is equation (\ref{35e}), which we prepare in a form
adapted to carrying out perturbation theory in powers of $\xi$,
\bea
\label{37a}
x^2 \vf '' + x \vf ' - ( x^2 + \nu ^2 ) \vf = \xi^2 \, x^{4-4k} \vf
\eea
The solutions for $\xi=0$ are the modified Bessel functions $I_\nu (x)$ and $K_\nu(x)$.
Regularity of the $\xi=0$ solution $\vf_0 $ near the horizon as $x \to \infty$
requires $\vf _0(x) = K_\nu (x)$ up to an overall normalization.  The Green function
corresponding to the differential operator on the left hand side of (\ref{37a})
with these large $x$ asymptotics  is given by,
\bea
\label{37bb}
G(x,y) = \theta (x-y) K_\nu (x) I_\nu (y) + \theta (y-x) K_\nu (y) I_\nu (x)
\eea
Using the Wronskian relation $x K_\nu I_\nu ' - x I_\nu  K'_\nu  =1$, it
 is readily checked that,
\bea
\label{37c}
\left ( x^2 \p_x^2  + x \p_x   - x^2 - \nu^2 \right ) G(x,y) = - y \delta (x-y)
\eea
The differential equation (\ref{37a}), together with the horizon asymptotics
may now be transformed into the following integral equation for $\vf$,
\bea
\label{37dd}
\vf (x) = K_\nu (x) - \xi ^2 \int _0 ^\infty dy \, G(x,y) \, y^{3-4k} \, \vf (y)
\eea
Next, we show that the perturbative solution to  (\ref{37dd}) is point-wise convergent
in the interval $1/2<k<1$.  The perturbative solution is obtained by iteration,
\bea
\label{A1}
\vf (x) =  \sum _{n=0}^\infty (- \xi^2 )^n \vf _n (x)
\hskip 1in
\vf _{n+1} (x) = \int _0 ^\infty dy \, G(x,y) \, y^{3-4k} \, \vf _n (y)
\eea
where $\vf _0(x) = K_\nu (x)$, and the Green function $G$ was given in (\ref{37bb}).
To investigate the convergence properties of the expansion in powers of $\xi$,
it will be useful to have the asymptotic behavior of the Bessel functions.
For $x \to 0$, we have,
\bea
\label{A3}
I_\nu (x)  \sim  \left ({x \over 2} \right )^\nu {1 \over \G(\nu+1)}
\hskip 1in
K_\nu (x)  \sim  \half \left ({x \over 2} \right ) ^{-\nu} \G(\nu)
\eea
while for  $x \to \infty$, we have,
\bea
\label{A4}
I_\nu (x) \sim  {e^x \over \sqrt{2 \pi x}}
\hskip 1in
K_\nu (x)  \sim  { e^{-x} \over \sqrt{2x/\pi}}
\eea
Subleading terms will not be needed here. We begin by considering the first
iteration,\footnote{The case $k > 1$ can be acommodated as well by relaxing the lower
integration boundary to a non-zero positive value (such as $\infty$),
and absorbing an infinite term proportional to $K_\nu(x)$ into $\vf_0(x)$.}

\bea
\label{A2}
\vf _1 (x) =
K_\nu (x)  \int _0 ^x dy \,  y^{3-4k} \, I_\nu (y) \, K_\nu (y)
+  I_\nu (x) \int _x ^\infty dy  \, y^{3-4k} \, K_\nu (y)^2
\eea
Clearly, for all $1/2<k<1$, both integrals are point-wise convergent for all finite $x$.

\sm

To investigate higher order terms in the expansion (\ref{A1}),
we need the asymptotic behavior of $\vf _1$, which may be obtained
from the asymptotics in (\ref{A3}) and (\ref{A4}).
For small $x$, the first integral in (\ref{A2})  behaves as $x^{4-4k}$, while the
second integral behaves as follows,
\bea
\int _x ^\infty dy  \, y^{3-4k} \, K_\nu (y)^2 & \sim &
\left \{ \matrix{x^{-2 \nu + 4 - 4k} & \hbox{if} & -2 \nu + 4  < 4k  \cr
x^0 & \hbox{if} & -2 \nu + 4  > 4k  \cr } \right .
\eea
Thus, as $x \to 0$, the behavior may be captured as follows,
\bea
\vf _1 (x)  \sim  x^\ep K_\nu(x) \qquad \ep = {\rm Min}(2 \nu, 4-4k)
\eea
which, for $k<1$, is softer than the leading order term $K_\nu(x)$.
For large $x$, the integrals behave as,
\bea
\int _0 ^x dy \,  y^{3-4k} \, I_\nu (y) \, K_\nu (y)  & \sim &
\left \{ \matrix{x^{3-4k} & \hbox{if} & k< 3/4  \cr x^0 & \hbox{if} & k > 3/4 \cr } \right .
\no \\
\int _x ^\infty dy  \, y^{3-4k} \, K_\nu (y)^2 ~~ & \sim & x^{2-4k} e^{-2x}
\eea
For $k > 1/2$, and as $x \to \infty$, the leading behavior arises entirely from the
first integral. For $k> 3/4$, the asymptotic behavior is precisely that of the leading
order term,
\bea
\vf _1 (x) \sim  K_\nu (x)
\eea
Thus, the perturbative expansion in this region of $k$ is uniformly convergent.
For $k<3/4$, an extra power arises, and we have,
\bea
\vf _1 (x) \sim  x^{3-4k} K_\nu (x)
\eea
the behavior of which is harder than the lowest order $K_\nu(x)$ term. Still,
the contribution to each order is point-wise convergent, and we easily derive
the following general leading asymptotic formula for $1/2<k<3/4$,
\bea
\vf _n (x) \sim {1 \over n!} \left ( { x^{3-4k} \over 2 (3-4k)} \right )^n K_\nu (x)
\eea
This leading order asymptotic behavior may be resummed, and we find,
\bea
\vf (x) \sim \exp \left ( - { \xi ^2 x^{3-4k} \over 2(3-4k)} \right ) K_\nu (x)
\eea
a formula which demonstrates that the perturbation theory is well-defined.

\subsection{WKB expansion}

A key observation throughout is that the powers of $p_\pm$ occurring in the expansion
parameters are independent of $\nu$. This suggests that, for certain regimes, we should
be able to take $\nu^2$ large and use WKB. To this end, we use equation
(\ref{35c}) as our starting point since it clearly exhibits both free parameters
$p^2$ and $p_-^2$. We now make the following change of variables,
\bea
\label{38a}
 \ln z = x + \ln \lambda \hskip 1in \ti \phi (z) = \vf (x)
\eea
where $\lambda$ is a constant. (The change of variables of (\ref{38a})
is not to be confused with the one of (\ref{35d}).)
The scalar equation (\ref{35c}) is transformed
into the standard  Schr\"odinger form,
\bea
\label{38bb}
- {d^2 \vf \over dx^2}
+ \left (  { p^2 \lambda ^2 \over b^3} \, e^{2x}
+ {p_-^2 a_M \lambda ^{4-4k} \over 36} \, e^{(4-4k)x} \right ) \vf
= - \nu^2 \vf
\eea
Although $p^2$ and $p_-^2$ appeared as two independent free parameters
in (\ref{35c}), only the combination $\xi$  of (\ref{35f}) is intrinsic, while $p$ merely
sets a scale for the coordinate $z$, or shifts an origin for the coordinate $x$.
The addition of the constant $\ln \lambda$ to $x$ precisely allows for such a shift.
We will choose $\lambda$ such that it sets the coefficients of both
potential terms in (\ref{38bb}) equal to one another, thereby shifting the natural
origin to $x=0$. The corresponding condition on $\lambda$ is as follows,
\bea
\label{38c}
\lambda ^{4k-2} = {  b^3 a_M  p_- \over 36 p_+}
\eea
With this value of $\lambda$, the common factor of the potential
terms becomes,
\bea
\label{38d}
{p^2 \lambda ^2 \over b^3}= {1 \over \hbar^2}
\hskip 1in
\hbar ^{4k-2} = { 1 \over \xi} =   {(p_+)^{2-2k} \over \xi_0 \, (p_-)^{2k}  }
\eea
and the Schr\"odinger equation of (\ref{38bb}) takes the form,
\bea
\label{38e}
- \hbar^2 {d^2 \vf \over dx^2}
+ \left (   e^{2x}  +  e^{(4-4k)x} \right ) \vf = - \hbar^2 \nu^2 \vf
\eea
Since the objects we are after will be independent of $\nu$, we take $\nu$ large.
Specifically, for $E=-\hbar ^2 \nu ^2 \gg 1$, this equation admits a perfectly fine
WKB expansion, provided $\hbar \ll 1$, or $\xi \gg 1$. This range of parameters
includes, for example, the long distance limit in which $p_+ \to 0$ at fixed $p_-$.
The WKB method involves a turning point at $x=x_*$, given by,
\bea
e^{2x_*} + e^{(4-4k)x_*} = E
\eea
For $x>x_*$, we have an exponentially decaying branch, while for $x<x_*$,
two oscillating exponentials contribute, whose behavior accounts for the $z^\nu$
and $z^{-\nu}$ asymptotic branches. The WKB method provides a systematic
expansion  in powers of $\hbar$, independently of the value of $E$.
Therefore, our results for the form of the expansion parameter should be
valid for all $E$, and thus all $\nu$, including the physical region
$\nu \geq 1$ of our problem.

\section{Linearized reduced equations in light-cone gauge}
\setcounter{equation}{0}
\label{appB}

The reduced fluctuation equations in light-cone gauge around the full
charged magnetic brane solutions were derived using Maple.
They are relatively involved, and will not be presented here.  In part,
they may be solved iteratively. The $--$ component of the Einstein
equation may be solved for $\hti_{rr}$ in terms of $\hti_K$ by,
\bea
\label{B1}
\hti_{rr} = - { 2  \over K_0 L_0^2} \hti_K
\eea
The $r-$ component of the Einstein equation may be solved for $\hti_{r+}$
in terms of $\hti_K$ by,
\bea
\label{B2}
\hti_{r+}= - {i \over p_-} \left ( { L_0 K_0'  \over  K_0^2 }\hti_K
+  { 2L_0  \over  K_0} \hti_K' \right )
\eea
The $-$ component of the Maxwell equation may be solved for $\ati_+$
in terms of $\ati_r$ by,
\bea
\label{B3}
\ati_+={i L_0^2 \over K_0 p_-} \left ( \left ( K_0L_0 \ati_r \right )' + 2kb \ati_r \right )
\eea
The $+-$ component of the Einstein equation may be solved for
$\hti_{++}$ in terms of $\hti_K$ and $\ati_r$ (the dependences of the other
fields having been eliminated using the above relations) by,
\bea
\label{B4}
\hti_{++}& = & - { L_0^2 \over 3 K_0^3 p_-^2}
\bigg ( -6 K_0^2 L_0^2 \hti_K'' -3 K_0K_0'L_0^2\hti_K' -12 K_0^2 L_0 L_0' \hti_K'
\no \\ && \hskip 0.8in
- 3 L_0^2 K_0 K_0'' \hti_K + 24 \hti_K + 3 L_0^2 (K_0')^2 \hti_K + 6 K_0^2 L_0 L_0'' \hti_K
\no \\ && \hskip 0.8in
-9 K_0 K_0' L_0L' _0 \hti_K + 6 K_0^2 (L_0')^2 \hti_K - 4i K_0^3 L_0 p_-A_0' \ati_r \bigg )
\eea
This leaves only the $r$ component of the Maxwell equation and the $rr$
component of the Einstein equation for the fields $\hti_K$ and $\ati_r$.
They satisfy the following pair of coupled second order equations,
\bea
\label{B5}
0 & = & \hti_K'' + { 2 L_0' \over L_0} \hti_K'- {K_0'\over K_0} \hti_K'
- { 8 \over L_0^2 K_0^2} \hti_K +{ 4 L_0' K_0' \over L_0K_0} \hti_K + { 2 K_0'' \over K_0} \hti_K
\no \\ &&
+ { M_0 p_-^2 \over L_0^4} \hti_K - { 2 p_+ p_- \over L_0^3} \hti_K -{ 8 \over L_0^2} \hti_K
-{ 2 (K_0')^2 \over K_0^2} \hti_K + { 8i K_0A_0' p_- \over 3L_0} \ati_r
\no \\
0 & = & - \half K_0L_0^4 \ati_r'' - 2 K_0L_0^3 L_0' \ati_r' - \half L_0^4 K_0' \ati_r'
-\half K_0L_0^3 L_0'' \ati_r - kb L_0^2 L_0' \ati_r
\no \\ &&
 -K_0L_0^2 (L_0')^2 \ati_r + 2kb {L_0^3 K_0' \over K_0} \ati_r
+  {L_0^4 (K_0')^2 \over 2K_0} \ati_r  + { 6 k^2 L_0^2 \over K_0} \ati_r - \half K_0M_0p_-^2 \ati_r
\no \\ &&
+p_+p_- K_0L_0 \ati_r -{3 \over 2} L_0^3 L_0' K_0' \ati_r - \half L_0^4 K_0'' \ati_r
+ i p_- L_0 A_0' \hti_K
\eea
The remaining equations are automatically obeyed.

\section{Gauge change in the overlap region: derivations}
\setcounter{equation}{0}
\label{appC}

In this Appendix, we present the derivation of the solutions in the overlap
region to the gauge change equations of (\ref{60e}), which we repeat here
for convenience,
\bea
\label{C1}
h_{MN} e^{ipx} & = & \hti_{MN} e^{ipx} + \nabla _M U_N + \nabla _N U_M
\no \\
a_M e^{ipx} & = & \ati_M e^{ipx} + U^K F_{KM} + \p_M \Theta
\eea
Imposing rotation invariance of (\ref{40b}) on $\hti_{MN}$, $\ati_M$,
$h_{MN}$, and $a_M$ gives $u^1=u^2=0$. We impose the conditions for
light-cone gauge of (\ref{40c}) on $\hti_{MN}$ and $ \ati_M$, and
for \ban gauge of (\ref{4e}) on $h_{MN}$, and $a_M$. We assign the
expressions for the light-cone gauge solution in the overlap region of (\ref{43a})
and (\ref{43d}) to the remaining components of  $\hti_{MN}$ and $\ati_M$.
Equations (\ref{C1}) for the remaining components of $h_{MN}$ are,
\bea
\label{C2}
h _{++} (r) & = & 4ib p_+ r u^-(r) +{8 i b q \over p_-} \left ( c_+ r^{2k} + c_- r \right )
- { 4 q^2 r^{2k-1} \over 2k-1} \, u^r(r) - { 4 i p_+ q^2 r^{2k} \over k(2k-1)} \, u^+(r)
\no \\
h _{+-} (r) & = & 2 b u^r(r)  + 2ib p_+ r u^+(r) + 2 i b p_- r u^-(r)
-{2i q^2 p_- r^{2k} \over k(2k-1)} \, u^+(r)
\no \\
h _{--} (r) & = & 4 i b p_- r u^+ (r)
\no \\
h _K (r) & = & v_+ r^\sigma + v_- r^{-1-\sigma}
\eea
while those for the gauge field are given  by,
\bea
\label{C3}
a_+ (r) & = & i p_+ \theta (r) +  (2k-1) {6 ib c_+ r^k \over p_-}  - q r^{k-1} u^r(r)
\no \\
a_- (r) & = &  i p_- \theta (r)
\eea
The equations determining $\theta $ and the remaining components of $u^M$
are given by,
\bea
\label{C4}
0 & = & u^r(r)' + i p_+ u^+(r) + i p_- u^- (r)
-{i q^2 p_- r^{2k-1} \over bk (2k-1)} \, u^+(r) - v_+ r^\sigma - v_- r^{-1-\sigma}
\no \\
0 & = & r u^+(r)' +{ ip_- u^r(r) \over 8b^3 r^2}
\no \\
0 & = & r u^-(r)' +{ ip_+ u^r(r) \over 8 b^3 r^2} -{q^2 r^{2k} \over bk (2k-1)} \, u^+(r)'
- {2 i  \over p_-} \left ( \sigma v_+ r^\sigma -(\sigma +1) v_- r^{-1-\sigma} \right )
\no \\
0 & = & \theta (r)'  +{c_+ \over 4} r^{k-2} +{c_- \over 4} r^{-1-k} + q r^{k-1} \, u^+(r)
\eea
From (\ref{C2}) and (\ref{C3}), it is clear that, in the overlap region,  the metric
and gauge fluctuation fields $h$ and $a$ are determined completely in terms
of $v_\pm$, $c_\pm$,  $u^r, u^\pm$, and $\theta $. In turn, $u^r, u^\pm$ and
$\theta$ are governed by differential equations
(\ref{C4}) which only involve the data $v_\pm, c_\pm$.

\subsection{Form of the \ban gauge solution}

Since the overlap region is contained in the far region, it is appropriate to match
the solutions in the overlap region  to (\ref{C2}), (\ref{C3}), and (\ref{C4}) with the
form of the \ban gauge solutions in the far region. The latter are already known
from equation (4.3) of \cite{D'Hoker:2010ij}, and may be  parametrized as follows,
\bea
\label{C5}
h _{++}(r) = s_{++} r + t_{++} & \hskip 1in & a_+ (r) = a_+ ^1 \, r^{+k} + a _+ ^0
\no \\
h _{--} (r) = s_{--} r + t_{--}  && a _- (r) = a_- ^1 \, r^{-k} + a _-^0
\no \\
h_{+-} (r) = s_{+-} r + t_{+-} &&
h_K \, (r) = v_+ r^\sigma + v_- r^{-\sigma -1}
\eea
where the coefficients $v_\pm, s_{\pm \pm}, t_{\pm \pm}, s_{+-}, t_{+-}, a^0_\pm$,
and $ a^1_\pm$ are independent of $r$, and the constraint equation implies $s_{+-}=0$.
Terms with different functional dependence on $r$ have been omitted from (\ref{C5}),
since they may be neglected in the overlap region.

\sm

From equations (\ref{C2}) and (\ref{C3}), we can infer which functional forms
of $u^r, u^\pm$, and $\theta$ are allowed to contribute to the functional forms
of $h_{\pm \pm}, h_{+-}, h_K$ and $a_\pm$ specified in (\ref{C5}).
The allowed functional forms are found to be as follows,
\bea
\label{C6}
u^+(r) & \hskip 0.6in & r^0, \, r^{-1}, \, r^{-2k}, \, r^{-2k+1}
\no \\
u^-(r) & & r^0, \, r^{-1}
\no \\
u^r(r) & & r^0, \, r^1, \, r^{-2k+1}, \, r^{-2k+2}
\no \\
\theta (r) && r^{+k}, \, r^{-k}
\eea
From these functional dependences, it is clear that all terms involving
$r^\sigma$ and $r^{-1-\sigma}$ may be dropped from (\ref{C4}), since they
can never contribute to the relevant functional forms of $u^r, u^\pm, \theta$
in the overlap region given in (\ref{C6}).

\sm

Furthermore, by scaling a factor $p^2=p_+p_-$ out of $r$, we see that several
terms in (\ref{C2}), (\ref{C3}), and (\ref{C4}) exhibit coefficients that contain
powers of $\xi \sim (p_+)^{k-1} (p_-)^k$ compared to other terms in the
same equation, and are thereby suppressed for $\xi \ll 1$.
This is the case for the fourth term of the second equation in (\ref{C2}),
and the fourth term in the first equation of (\ref{C4}), both of which
are suppressed by $\xi^2$ compared to the second terms in each equation.
It is also the case for the third term in the third equation of (\ref{C4}) which is
suppressed by $\xi^2$ compared to the second term, upon making use of
the second equation of (\ref{C4}).

\sm

Omitting all irrelevant terms leaves simplified equations for the
metric fluctuations,
\bea
\label{C7}
h _{++} (r) & = & 4ib p_+ r u^-(r) +{8 i b q \over p_-} \left ( c_+ r^{2k} + c_- r \right )
- { 4 q^2 r^{2k-1} \over 2k-1} \, u^r(r) - { 4 i p_+ q^2 r^{2k} \over k(2k-1)} \, u^+(r)
\no \\
h _{+-} (r) & = & 2 b u^r(r)  + 2ib p_+ r u^+(r) + 2 i b p_- r u^-(r)
\no \\
h _{--} (r) & = & 4 i b p_- r u^+ (r)
\no \\
h _K (r) & = & v_+ r^\sigma + v_- r^{-\sigma -1}
\eea
and for the gauge field fluctuations,
\bea
\label{C8}
0 & = & u^r(r)' + i p_+ u^+(r) + i p_- u^- (r)
\no \\
0 & = & r u^+(r)' +{ ip_- u^r(r) \over 8b^3 r^2}
\no \\
0 & = & r u^-(r)' +{ ip_+ u^r(r) \over 8 b^3 r^2}
\no \\
0 & = & \theta (r)'  +{c_+ \over 4} r^{k-2} +{c_- \over 4} r^{-1-k} + q r^{k-1} \, u^+(r)
\eea
Note that the fluctuation $h_K$ decouples from all the others.

\subsection{Solutions in \ban gauge for metric fluctuations}

To solve the simplified equations of (\ref{C7}) and (\ref{C8}), we begin by taking the
derivative of the first equation in (\ref{C8}), and eliminating the derivatives of
$u^\pm$ using the second and third equations. This gives
the following equation for $u^r$ alone,
\bea
\label{C9}
0= u^r(r)''  + {p^2 \over 4b^3 r^3} u^r(r)
\eea
The second term may be neglected in the overlap region, since it is
suppressed by a factor of $p^2/r \ll 1$. We are left with $u^r(r)''=0$. The
solutions to this equation and to the second and third equations in (\ref{C8})
may be parametrized as follows,
\bea
\label{C10}
u^r (r) & = & u^r_1 r + u^r _0
\no \\
u^+(r) & = & u^+_0 +{ i p_- u^r_1 \over 8 b^3 r}
\no \\
u^-(r) & = & u^-_0 +{ i p_+ u^r_1 \over 8 b^3 r}
\eea
The integration constants $u^\pm _0, u^r_0, u^r_1$ are independent of $r$.
In the result for $u^\pm$ an extra term proportional to $u_0^r/r^2$ has been omitted,
since its functional dependence is not in the list of (\ref{C6}).
Enforcing also the first equation of (\ref{C8}), and omitting a term suppressed by
a factor of $p^2/r$, gives the following non-trivial relation between the integration constants,
\bea
\label{C11}
u_1^r =- i p_+ u^+_0 - i p_- u^-_0
\eea
We note that the functional dependences involving  $r^{-2k}$, which were listed in
(\ref{C6}) for $u^+, u^r$, do not actually occur. As a result, the last
two terms of $h_{++}$ in (\ref{C7}) do not contribute.  The
term in $c_+$ will also not contribute there because of its functional dependence,
and neither will the term in $c_-$ because it is momentum-suppressed
in view of (\ref{43c}). As a result, the metric fluctuations are indeed given by
(\ref{C5}) and (\ref{61a}), with the coefficients given in (\ref{61b}).

\subsection{Solution in \ban gauge for gauge field fluctuations}

Using the above result for $u^+$, the solution for $\theta $ in the overlap region
may be readily computed, and we have,
\bea
\label{C13}
\theta (r) = \theta _0 - {q u_0^+ \over k} \, r^k +  {c_- \over 4k} \, r^{-k}
\eea
where $\theta _0$ is a further integration constant. Using this expression
in the solutions for $a_\pm$ of (\ref{C3}), and retaining only the functional
behaviors of the solutions (\ref{C5}) suitable in the overlap region, we find,
\bea
a _+ ^0 = ip_+ \theta _0 & \hskip .8in &
a_+^1 =  (2k-1) {6ib c_+ \over p_-}  + {qp_+ (k-1) \over 4b p_- k}  s_{--}
+ { q p_- \over 4  b p_+} s_{++}
\no \\
a _- ^0 =  i p_- \theta _0 && a_- ^1 =  {i p_- c_-  \over 4k}
\eea
Within our low momentum approximation, we may further set $c_-=0$.

\end{document}